\documentclass[%
 aip,
 amsmath,amssymb,
 longbibliography,
 reprint,%
]{revtex4-1}

\usepackage{graphicx}% Include figure files
\usepackage{bm}% bold math
\usepackage[utf8]{inputenc}
\usepackage[T1]{fontenc}
\usepackage{mathptmx}
\usepackage{xcolor}
\usepackage{soul}
\usepackage{color}
\usepackage{multirow}
\usepackage{xspace}
\usepackage{amsmath}
\usepackage{amsfonts}
\usepackage{amssymb}
\usepackage{amsthm,bm}
\usepackage{commath}
\usepackage{setspace}
\usepackage{titlesec}

\titleformat{\paragraph}[runin]
  {\normalfont\normalsize\bfseries\itshape}{(\theparagraph)}{1em}{}
\titlespacing*{\paragraph}
  {0pt}{*}{0.5em}

\titleformat{\subparagraph}[runin]
  {\normalfont\normalsize\itshape}{(\thesubparagraph)}{1em}{}
\titlespacing*{\subparagraph}
  {0pt}{*}{0.5em}

\renewcommand\theparagraph{\roman{paragraph}}
\renewcommand\thesubparagraph{\alph{subparagraph}}

\makeatother

\newcommand{\D}{^{\dagger}}

\newcommand{\bd}{{\bf d}}
\newcommand{\hbd}{\hat{\bf d}}
\newcommand{\br}{{\bf r}}
\newcommand{\hbr}{\hat{\bf r}}

\newcommand{\bE}{{\bf E}}
\newcommand{\hbE}{\hat{\bf E}}
\newcommand{\calF}{\mathcal{F}}

\newcommand{\cH}{\hat{H}}

\newcommand{\be}{\begin{equation}}
\newcommand{\ee}{\end{equation}}
\newcommand{\bea}{\begin{eqnarray}}
\newcommand{\eea}{\end{eqnarray}}
\newcommand{\beqa}{\begin{eqnarray*}}
\newcommand{\eeqa}{\end{eqnarray*}}
\newcommand{\nn}{\nonumber}

\newcommand{\ba}{\begin{array}{c}}
\newcommand{\baa}{\begin{array}{cc}}
\newcommand{\baaa}{\begin{array}{ccc}}
\newcommand{\baaaa}{\begin{array}{cccc}}
\newcommand{\ea}{\end{array}}

\newcommand{\bma}{\left[\begin{array}{c}}
\newcommand{\bmaa}{\left[\begin{array}{cc}}
\newcommand{\bmaaa}{\left[\begin{array}{ccc}}
\newcommand{\bmaaaa}{\left[\begin{array}{cccc}}
\newcommand{\ema}{\end{array}\right]}

\begin{document}

\preprint{AIP/123-QED}

\title{Tunable giant Purcell enhancement of quantum light emitters by means of acoustic graphene plasmons}
% Force line breaks with \\
\author{Justin Gruber}
\affiliation{College of Optics and Photonics (CREOL), University of Central Florida, Orlando, FL 32826, USA}
\affiliation{The Institute of Optics, University of Rochester, Rochester, NY 14627, USA}
\author{Mahtab A. Khan}
\affiliation{NanoScience Technology Center, University of Central Florida, Orlando, FL 32826, USA.}
\affiliation{Department of Physics, Federal Urdu University of Arts, Sciences and Technology, Islamabad, Pakistan.}
\author{Dirk R. Englund}
\affiliation{Department of Electrical Engineering and Computer Science, Massachusetts Institute of Technology, Cambridge, MA 02139, USA}
\author{Michael N. Leuenberger}
\affiliation{NanoScience Technology Center, Department of Physics, College of Optics and Photonics, University of Central Florida, Orlando, Fl 32826, USA.}
\email{michael.leuenberger@ucf.edu}

%\affiliation[2]{ 
%NanoScience Technology Center, Department of Physics, and College of Optics and Photonics, University
%of Central Florida, Orlando, FL 32826, USA
%\\This line break forced with \textbackslash\textbackslash
%email: michael.leuenberger@ucf.edu
%}%

%\date{\today}% It is always \today, today,
             %  but any date may be explicitly specified

\begin{abstract}
%It has long been known that plasmons can be used to cause fluorescence enhancement: the increase of the emission rate of a quantum emitter through an external electronic field. This effect has been shown using surface plasmons in metallic nanoparticles as well as in graphene through graphene surface plasmons (GSPs) or localized surface plasmons (LSPs). 
%Here, we propose the use of acoustic graphene plasmons (AGPs) to achieve tunable giant Purcell enhancement factors of single photon, entangled-photon, and multipolar quantum emitters in 2D materials. 
Inspired by the remarkable ability of plasmons to boost radiative emission rates, we propose leveraging acoustic graphene plasmons (AGPs) to realize tunable, giant Purcell enhancements for single-photon, entangled-photon, and multipolar quantum emitters. These AGPs are localized inside a cavity defined by a graphene sheet and a metallic nanocube and filled with a dielectric of thickness of a few nanometers and consisting of stacked layers of 2D materials, containing impurities or defects that act as quantum light emitters. Through finite-difference time domain (FDTD) calculations, we show that this geometry can achieve giant Purcell enhancement factors over a large portion of the infrared (IR) spectrum, up to 6 orders of magnitude in the mid-IR and up to 4 orders of magnitude at telecommunications wavelengths, reaching quantum efficiencies of 95\% and 89\%, respectively, with high-mobility graphene. We obtain Purcell enhancement factors for single-photon electric dipole (E1), electric quadrupole (E2), and electric octupole (E3) transitions and two-photon spontaneous emission (2PSE) transitions, of the orders of $10^{4}$, $10^{7}$, $10^{9}$, and $10^9$, respectively, and a quantum efficiency of 79\% for entangled-photon emission with high-mobility graphene at a wavelength of $\lambda=1.55$ $\mu$m. Importantly, AGP mode frequencies depend on the graphene Fermi energy, which can be tuned via electrostatic gating to modulate fluorescence enhancement in real time. As an example, we consider the Purcell enhancement of spontaneous single- and two-photon emissions from an erbium atom inside single-layer (SL) WS$_2$. Our results could be useful for electrically tunable quantum emitter devices with applications in quantum communication and quantum information processing.
\end{abstract}


\maketitle
\begin{figure*}[htb]
    \centering
    \includegraphics[width=\textwidth]{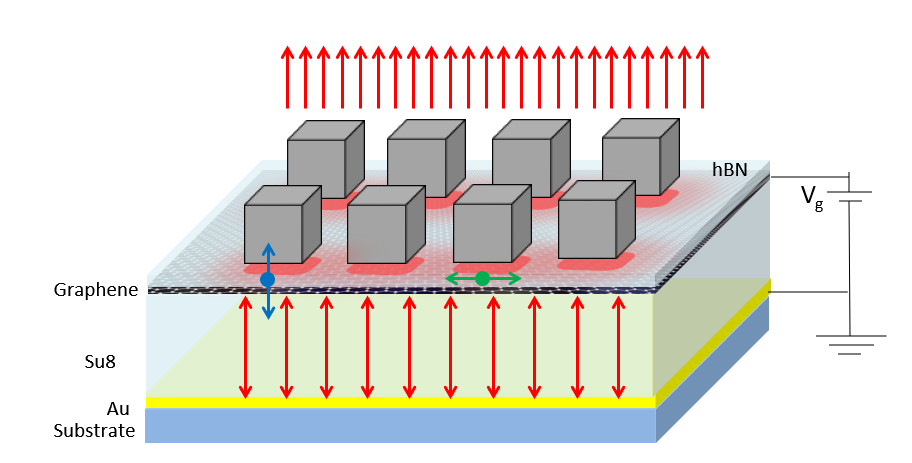}
    \caption{Schematic of proposed AGP cavity and heterostructure. Dipole emitter is located within the hBN layer, with the perpendicular orientation shown in blue and the parallel orientation shown in green. The acoustic graphene plasmons (AGPs) are localized in the region between the nanocubes and the graphene layer(s).}
    \label{fig:cavity}
\end{figure*}

\par
Single-photon emitters (SPEs) are a crucial technology in the development of quantum information technologies, and several methods for creating solid-state SPEs have been put forth.\cite{Aharonovich2016} Methods for creating SPEs include semiconductor quantum dots\cite{Senellart_Solomon_White_2017} and color centers caused by defects such as nitrogen-vacancy centers in diamond.\cite{https://doi.org/10.1002/adom.201400189} An emerging class of SPEs are defect states in two-dimensional (2D) materials. Single photon emission has been demonstrated experimentally in a variety of 2D materials including hexagonal boron nitride (hBN),\cite{Tran_Bray_Ford_Toth_Aharonovich_2015} transition metal dichalcogenides (TMDs),\cite{Tonndorf2015,Srivastava2015,Chakraborty2015,Koperski2015,He2015,Dass2019} and GaSe.\cite{Tonndorf_2017} 

Silicon has recently emerged as a promising host for telecom-band single-photon sources based on color centers, compatible with foundry-scale photonics. First isolations of individual carbon-related G centers in silicon-on-insulator (SOI) established bright, linearly polarized emission near 1.28~$\mu$m,\cite{Redjem2020NatElec} followed by surveys revealing multiple defect families and near-infrared single-photon emission~\cite{Durand2021PRL} and by wafer-scale, position-controlled creation of single G and W centers.\cite{Hollenbach2022NatComm} Foundational studies clarified photophysics and localization effects~\cite{Ivanov2022PRB} and the subtle distinction between genuine and faux single G centers in carbon-implanted Si,\cite{Durand2024PRB} while measurements resolved fine-structure and dipole characteristics of individual G centers in SOI structures.\cite{Durand2024PRX} On the device side, Englund's group demonstrated individually addressable, waveguide-coupled G centers with spectral trimming on SOI photonic integrated circuits;\cite{Prabhu2023NatComm} extended this to tunable, foundry-scale circuits;\cite{Larocque2024NatComm} and together with collaborators showed indistinguishable photon generation from a silicon artificial atom directly in silicon photonics.\cite{Komza2024NatComm} Complementing this, the Kante/Qarony collaboration reported the first all-silicon atom–cavity source by embedding a single G center in a silicon photonic-crystal cavity, achieving strong Purcell acceleration and near-unity coupling efficiency.\cite{Redjem2023NatComm} For the T center---a carbon–hydrogen complex attractive for spin–photon interfacing---high-efficiency single-photon emission was realized using a nanobeam geometry,\cite{Lee2023ACSPho} and cavity-QED coupling of single T centers has now been demonstrated, with rate enhancement and improved zero-phonon-line (ZPL) extraction.\cite{Islam2023,Johnston2024NatComm} In parallel, the Reiserer group established cavity-enhanced single-photon emission in silicon using nanophotonic resonators (including spin-resolved spectroscopy and Purcell factors of 78)~\cite{Gritsch2023Optica} and later achieved cavity-enhanced single artificial atoms at telecommunications wavelengths.\cite{Saggio2024NatComm} Together, these advances mark rapid progress from discovery to engineering: deterministic, CMOS-compatible placement,\cite{Hollenbach2022NatComm} active spectral control and indistinguishability on-chip,\cite{Prabhu2023NatComm,Komza2024NatComm,Larocque2024NatComm} and cavity integration across both G and T centers.\cite{Redjem2023NatComm,Islam2023,Johnston2024NatComm,Saggio2024NatComm,Gritsch2023Optica}

Er$^{3+}$ centers provide quantum emission directly in the low-loss telecom C-band, making them natural building blocks for fiber-connected quantum networks. In silicon nanophotonics, narrow optical transitions~\cite{Gritsch2022PRX} and strong Purcell enhancement in nanobeam cavities~\cite{Gritsch2023Optica} have enabled cavity-coupled single-ion interfaces with optical single-shot spin readout~\cite{Gritsch2025NatComm} and long optical/spin coherence in isotopically enriched $^{28}$Si.\cite{Berkman2025npjQI} These advances build on the earliest single-ion addressing of Er in Si~\cite{Yin2013Nature} and recent demonstrations of reliable foundry integration of Er in low-loss silicon waveguides.\cite{Rinner2023Nanophotonics} Beyond Si, Awschalom and co-workers established a CMOS-compatible thin-film platform based on Er:TiO$_2$ on silicon, showing large Purcell factors and on-chip integration,\cite{Dibos2022NanoLett} followed by direct isolation of individual Er emitters in TiO$_2$ films~\cite{APL2024TiO2} and systematic optical/microstructural studies.\cite{Singh2024JAP} In non-silicon crystals engineered for low spectral diffusion, indistinguishable single photons from a single Er$^{3+}$ in CaWO$_4$ coupled to silicon nanophotonics have been observed with high Hong–Ou–Mandel visibility,\cite{Ourari2023} and spin–photon entanglement from a single Er ion has been reported.\cite{Uysal2025PRX} Additional control knobs include electro-optic tuning and emission-rate control of single Er in thin-film LiNbO$_3$ photonic crystal cavities,\cite{Yang2023NatComm,Yang2021OE} and Stark tuning of single Er in Y$_2$SiO$_5$ by on-chip electrodes.\cite{Huang2023CPL} Together, these results map a clear route from discovery to engineering: telecom-band single photons and spin–photon interfaces in silicon,\cite{Gritsch2022PRX,Gritsch2023Optica,Gritsch2025NatComm,Rinner2023Nanophotonics,Berkman2025npjQI} scalable thin-film TiO$_2$ photonics,\cite{Dibos2022NanoLett,APL2024TiO2,Singh2024JAP} and host crystals tailored for indistinguishability and entanglement.\cite{Ourari2023,Uysal2025PRX,Huang2023CPL}

Two-dimensional (2D) semiconductors and insulators have rapidly emerged as versatile platforms for on-chip quantum light sources. In TMDs, spatially localized excitons in monolayer WSe$_2$ were the first 2D SPEs demonstrated in 2015, showing antibunched emission from defect/strain-localized states and inaugurating the field of quantum light in van der Waals materials.\cite{He2015,Srivastava2015,Chakraborty2015,Koperski2015,Tonndorf2015} Since then, deterministic nano-/micro-strain engineering has enabled spatially ordered arrays of TMD SPEs and controlled polarization selection rules,\cite{Branny2017,Paralikis2024} while plasmonic and dielectric nanocavities have provided Purcell enhancement and improved out-coupling for integrated photonics.\cite{Luo2018,So2021} Beyond optical pumping, electrically driven devices based on WSe$_2$ tunneling junctions now realize site-controlled and even array-level single-photon electroluminescence.\cite{So2021,Guo2023,Howarth2024}

A second major platform is hexagonal boron nitride (hBN), whose deep-level point defects host bright and photostable SPEs that operate at room temperature across visible to ultraviolet wavelengths.\cite{Tran2016,Bourrellier2016,Grosso2017,Shotan2016,Noh2018,Proscia2018,Patel2022,Sajid2020Review,Aharonovich2022MiniReview,Prasad2024,Mendelson2020} Considerable progress has been made in understanding spectral inhomogeneity, strain and Stark tuning, and in the near-deterministic activation/placement of emitters compatible with wafer-level processing and photonic integration.\cite{Mendelson2020,Proscia2018,Noh2018,Im2025,Li2022UV,Chen2025,Prasad2024} Together with robust room-temperature operation, these advances position hBN as a leading candidate for scalable quantum photonic circuits.

Recently, moiré superlattices in twisted TMD heterobilayers introduced a third route to single-photon emission via moiré-trapped interlayer excitons (IXs). These emitters exhibit discrete, highly tunable lines, long lifetimes, and access to spin/valley physics, and they have been interfaced with high-$Q$ nanocavities.\cite{Baek2020,Qian2024,Wang2024,Alexeev2024,Cai2023,Forg2021,BrotonsGisbert2021} Beyond fundamental interest, moiré IXs offer device-level knobs (twist angle, reconstruction, gating) that complement strain/defect control in monolayers.\cite{Lin2024} 

Across all 2D platforms, the state-of-the-art has moved from proof-of-concept to engineering: room-temperature operation (hBN), deterministic positioning (strain-patterned TMDs and processed hBN), electrical injection (WSe$_2$ diodes), polarization/valley control, and cavity integration for enhanced brightness and indistinguishability.\cite{Aharonovich2022MiniReview,Zhang2024Review,So2021} Remaining challenges include identifying and stabilizing microscopic defect configurations (hBN), mitigating spectral diffusion and dephasing, achieving narrow linewidths and high photon indistinguishability, and unifying electrical control with large-scale, CMOS-compatible integration.\cite{Sajid2020Review,Aharonovich2022MiniReview,Im2025,Chen2025}

Additionally, telecommunications makes extensive use of light at wavelengths near 1550 nm for transmission in fiber optics with minimum losses.\cite{Bonneville:24, 4810183} For this reason, it has been of great interest to extend the emission range of SPEs to include 1550 nm. This has been achieved through the doping of TMDs, such as MoS$_2$ and WSe$_2$  with erbium (Er) ions.\cite{https://doi.org/10.1002/adma.201601833,HUANG2021128610} Alternatively, the engineering of defects in hBN has been proposed as a potential SPE near 1550 nm. \cite{nano12142427}
\par
In order to improve the performance of an SPE, it can be placed inside an optical cavity, which increases the emission rate $\Gamma$ of the SPE. If we take $\Gamma _0$ to be the emission rate in vacuum and $\Gamma_{dip}$ to be the emission rate in the cavity, then the ratio $\calF=\frac{\Gamma_{dip}}{\Gamma _0}$ is known as the Purcell factor of the SPE-cavity system. This value can be related to the power radiated by the dipole in vacuum, $P _0$, and the power radiated by the dipole in the cavity, $P_{dip}$, where $\calF=\frac{\Gamma_{dip}}{\Gamma _0} = \frac{P_{dip}}{P _0}$. This $P_{dip}$ is made up of two components, the power that is radiated and the power that is lost due to nonradiative decay, denoted by $P_{rad}$ and $P_{loss}$, respectively. We can thus define the quantum efficiency (QE) of the system to be $QE = \frac{P_{rad}}{P_{rad}+P_{loss}}$. If we are only interested in the power that is radiated to the far field, we can define the radiative enhancement as $\frac{P_{rad}}{P_{0}}$. This value quantifies the increase in radiated light that is able to be collected in the far field, normalized to the amount of far field light radiated by the SPE in vacuum.
\par
It was first shown by Koppens and collaborators that it is possible to achieve strong coupling between a graphene nanodisk and an SPE.\cite{Koppens_Chang_García_de_Abajo_2011} The possibilities of engineering light-matter interaction via graphene nanodisks has been further explored including demonstration of Purcell enhancement of quadrupole transitions,\cite{Sanders_May_Alabastri_Manjavacas_2018} and a plasmon-plasmon interaction based the plasmon blockade effect.\cite{Manjavacas_Nordlander_García_de_Abajo_2012} It has also been shown by Cox and collaborators that the dipole-dipole interaction between a quantum well and a graphene nanodisk can be tuned via a nonlinear photonic crystal.\cite{Cox_Singh_Gumbs_Anton_Carreno_2012a} Alternative graphene geometries have shown promise in facilitating strong light-matter interaction as well, including a half-space graphene sheet which exhibits Purcell factors of up to 6 orders of magnitude \cite{Karanikolas_Tozman_Paspalakis_2019} and nanoislands which can strongly enhance multipolar transitions to rates exceeding the dipole transition rate.\cite{Rosolen_Maes_2021} In all cases presented above, the fluorescence enhancement is achieved by means of coupling to graphene surface plasmons (GSPs).
\par
There have been several demonstrations of the Purcell enhancement of emission from defect-based emitters in hBN by means of plasmonic nanoantennas. These include multi-photon effects with intensity enhancements as large as 250,\cite{https://doi.org/10.1002/adma.202106046,doi:10.1021/acs.nanolett.2c03100} and single photon emission with intensity enhancements of 200\%.\cite{https://doi.org/10.1002/adom.202300392} These emitters make use of the negatively charged boron vacancy ($V_B$) in hBN, which has spin-dependent transitions \cite{Gottscholl2020-ox} that can be controlled at room temperature.\cite{doi:10.1126/sciadv.abf3630} These room-temperature spin-dependent transitions make this SPE a promising candidate for use in quantum technologies based on spin states \cite{Wolfowicz2021-on,Awschalom2018-ws}.
\par
%In this work, we propose the enhancement of the fluorescence rate in SPEs in 2D materials via coupling to AGPs, rather than GSPs in graphene sheets or LSPs in nanoantennas.  These AGPs are supported in a cavity consisting of one or more 2D materials between a graphene sheet and a metallic nanoparticle, such as a silver (Ag) nanocube. The use of AGPs is attractive, as they offer a larger confinement factor than do GSPs due to the extreme localization within the 2D material.\cite{doi:10.1126/science.abb1570} We predict a strong Purcell enhancement of the fluorescence rate due to the large confinement of the AGPs in this structure. Importantly, we predict that the frequency at which this Purcell enhancement occurs can be  tuned by means of a gate voltage applied to the graphene sheet. This allows for the enhancement at a given wavelength to be switched on and off by shifting the resonance wavelength of the AGPs, with an on-off ratio of 303 (or 25.4 dB) at telecom wavelengths

Motivated by the exceptional ability of surface plasmons to boost radiative emission rates, we propose using \emph{acoustic graphene plasmons} (AGPs) to realize \emph{electrically tunable, giant} Purcell enhancements for single-photon, entangled-photon, and multipolar quantum emitters. Unlike graphene surface plasmons (GSPs) in extended sheets or localized surface plasmons (LSPs) in nanoantennas, AGPs concentrate the optical field in a nanogap cavity formed by a graphene sheet and a metallic nanocube, with a few-nanometer-thick dielectric spacer composed of stacked 2D layers that host defect/impurity-based quantum emitters.\cite{doi:10.1126/science.abb1570} Our finite-difference time-domain (FDTD) simulations show \emph{giant} Purcell factors over a broad infrared window, reaching up to $10^{6}$ in the mid-IR and up to $10^{4}$ at telecommunications wavelengths, with corresponding quantum efficiencies as high as $95\%$ (mid-IR) and $89\%$ (telecom) for high-mobility graphene. At $\lambda=1.55~\mu$m, we find Purcell enhancements of order $10^{4}$ for electric-dipole (E1) transitions and up to $10^{7}$ and $10^{9}$ for electric-quadrupole (E2) and electric-octupole (E3) transitions, respectively (see Table~\ref{tab:enhancement_factors} and Sec.~VII in SI); remarkably, two-photon spontaneous emission (2PSE) transitions also reach $\sim10^{9}$. Crucially, the AGP resonance is governed by the graphene Fermi energy and can thus be modulated via gate voltage, enabling real-time on–off control of the fluorescence enhancement (with large on–off ratios at telecommunications wavelengths) by shifting the AGP mode into or out of resonance with the emitter. As a concrete example, we analyze single- and two-photon spontaneous emission from Er$^{3+}$ centers embedded in single-layer WS$_2$, demonstrating that AGP cavities provide a compact, CMOS-compatible route to high-brightness, high-efficiency, and \emph{electrically tunable} quantum-light sources for quantum communication and information processing.

\begin{figure*}[htb]
    \centering
    \includegraphics[width=0.9\textwidth]{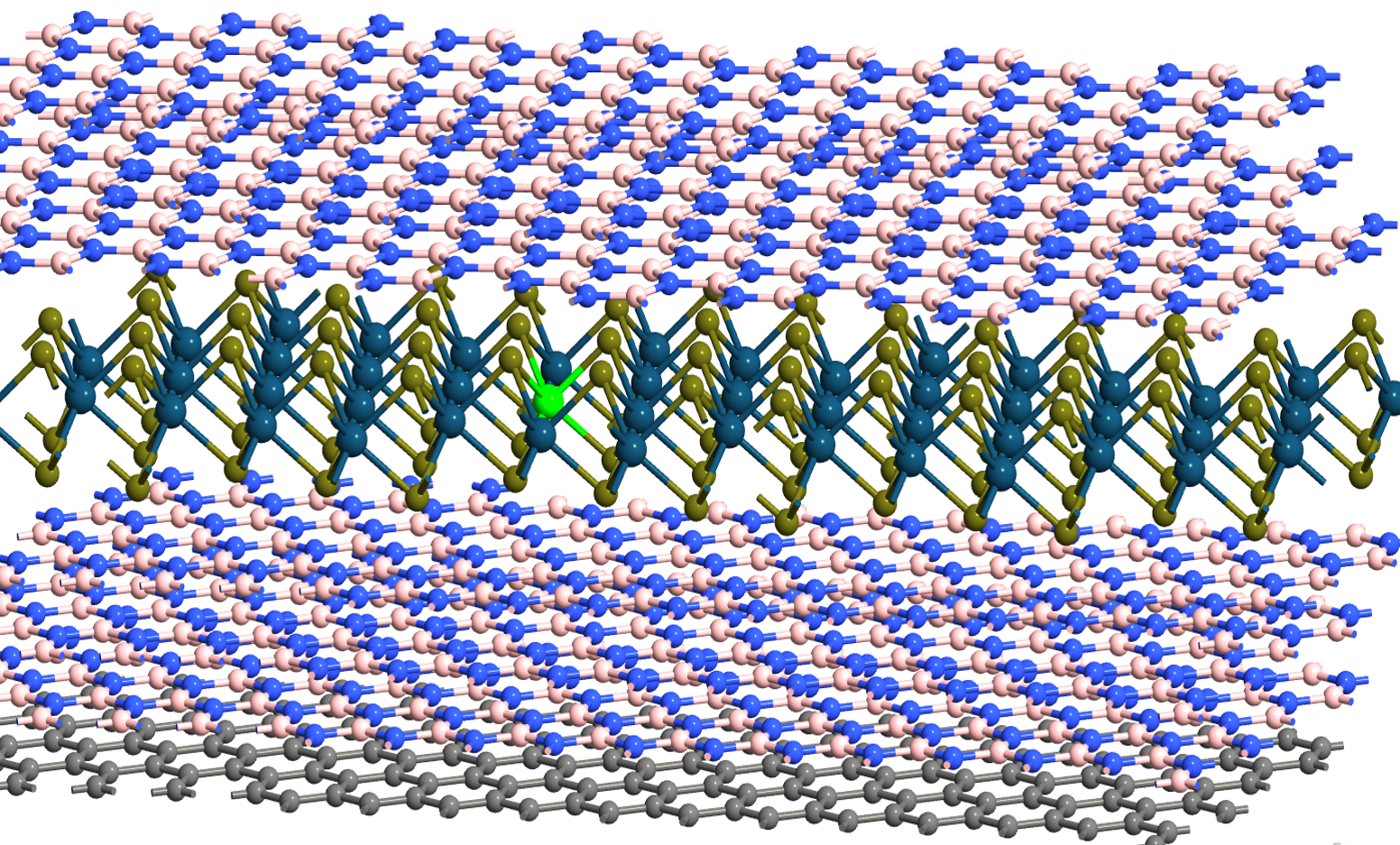}
    \caption{Schematic diagram showing Er (green ball) doping in tungsten (dark blue ball) disulphide (mustard color ball). A layer of SL WS$_2$ is sandwiched between layers of hexagonal BN (pink and blue balls). Graphene layer (grey balls) is also shown at the bottom.}
    \label{fig:schematic_Er_doping}
\end{figure*}

\section{Results and discussion}
\subsection*{AGP Resonances in the Heterostructure}
 The architecture of the AGP resonator is depicted in Fig.~\ref{fig:cavity}. We use an Ag nanocube with a side length of 50 nm placed upon one or more layers of hBN, a monolayer of Er-doped WS$_2$, and a graphene sheet. Underneath the graphene monolayer is a 1400 nm spacer made of SU-8 and an Au back mirror. The AGPs are excited in the hBN/WS$_2$/hBN heterostructure (see Fig.~\ref{fig:schematic_Er_doping}) between the graphene sheet and the metal nanoparticle. The thickness $h$ of the hBN/WS$_2$/hBN heterostructure, the number of graphene layers $C_S$ with intercalated FeCl$_3$, and the Fermi energy $E_F$ of the graphene layers were varied to tune the wavelength of the AGPs. 
\par
These AGPs induce an intense localized electrical field within the hBN/WS$_2$/hBN heterostructure. Here, we show that a single-photon emitter (SPE) in the form of erbium in WS$_2$ experiences a significant fluorescence enhancement effect due to the presence of these strong electric fields. These fields exist at the specific frequencies of the permitted AGP eigenmodes in the cavity. Consequently, fluorescence enhancement will selectively occur at these same frequencies. The AGP eigenmode frequencies of this cavity geometry have been predicted to follow the equation\cite{Craft2023}
\begin{equation}
  f_{mn}=\sqrt{\alpha \frac{chE_F}{\pi^2\hbar \epsilon_r}} \sqrt{\left(\frac{m \pi}{L}\right)^2 + \left(\frac{n \pi}{W}\right)^2}
  \label{eq:AGP_eigenfrequency}
\end{equation}
where $\alpha$ is the fine structure constant and $\epsilon_r$ is the relative permittivity of the hBN/WS$_2$/hBN heterostructure. The variables $L$ and $W$ are the length and width of the cavity, respectively, which are defined by the size of the Ag nanocube. 
\par
To characterize the fluorescence enhancement induced by these AGPs, we simulate this device using the Ansys Lumerical finite-difference time-domain (FDTD) software. We model the SPE as an electric dipole located within the WS$_2$ monolayer. We study two cases: a dipole oriented perpendicular to the the graphene sheet, and a dipole oriented parallel to the graphene sheet.
\par
\begin{figure*}
    \centering
    \includegraphics[ width=\textwidth]{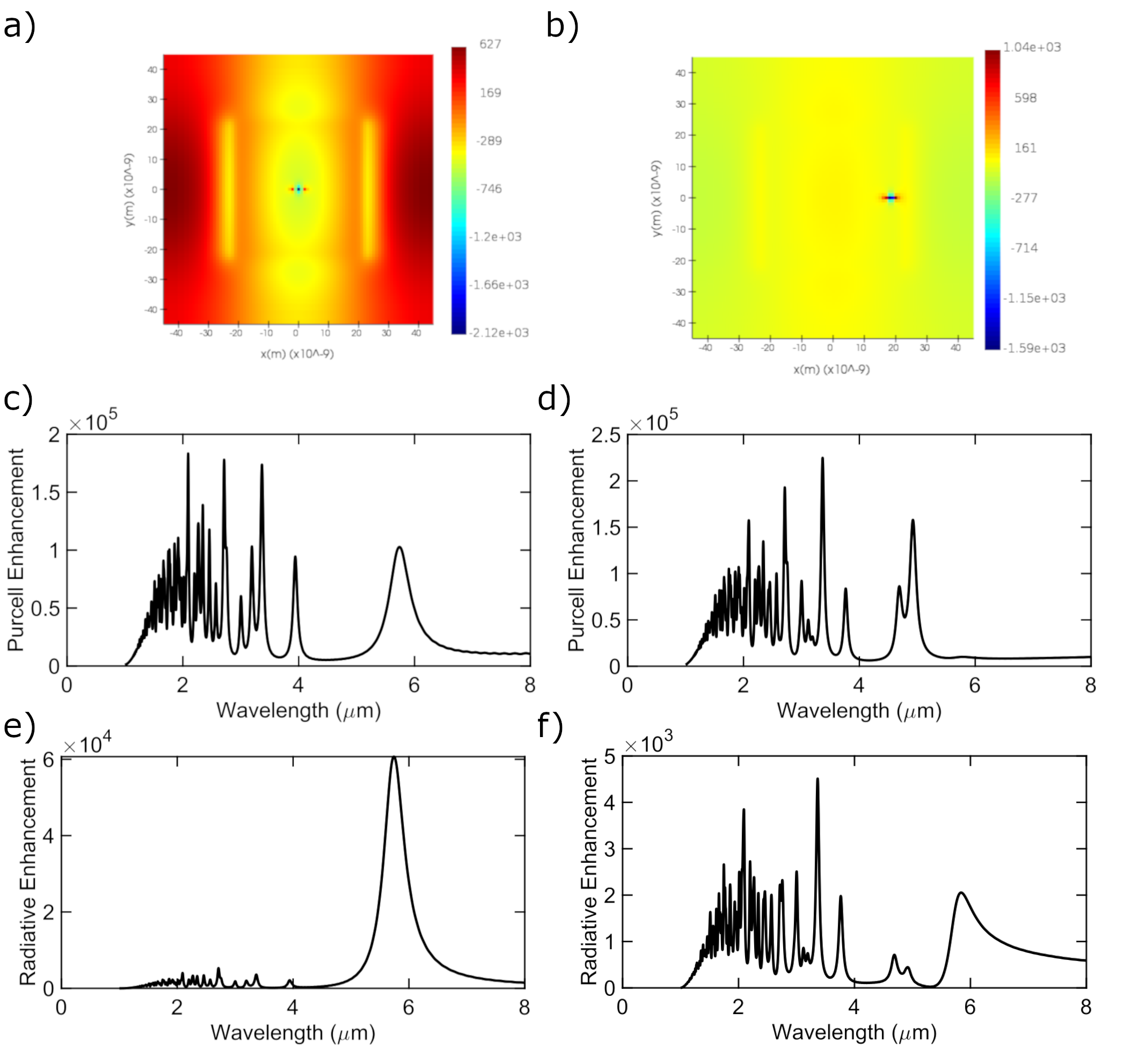}
    \caption{$E_X$ field of the AGP at the center of the heterostructure cavity with respect to the z-axis for the case when the dipole is oriented parallel to the graphene layer and placed both a) on resonance at x = 0 nm and b) off resonance at x = 18 nm. A large Purcell enhancement is seen c) on resonance and also d) off resonance. We see that the radiative enhancement e) on resonance is significantly larger than f) off resonance as well. This is because when the dipole is on resonance, it is located near the maximum value of $E_X$, whereas when off resonance it is located near a local minima of $E_X$.}
    \label{fig:ParallelOnOff}
\end{figure*}
\begin{figure*}
    \centering
    \includegraphics[width=\textwidth]{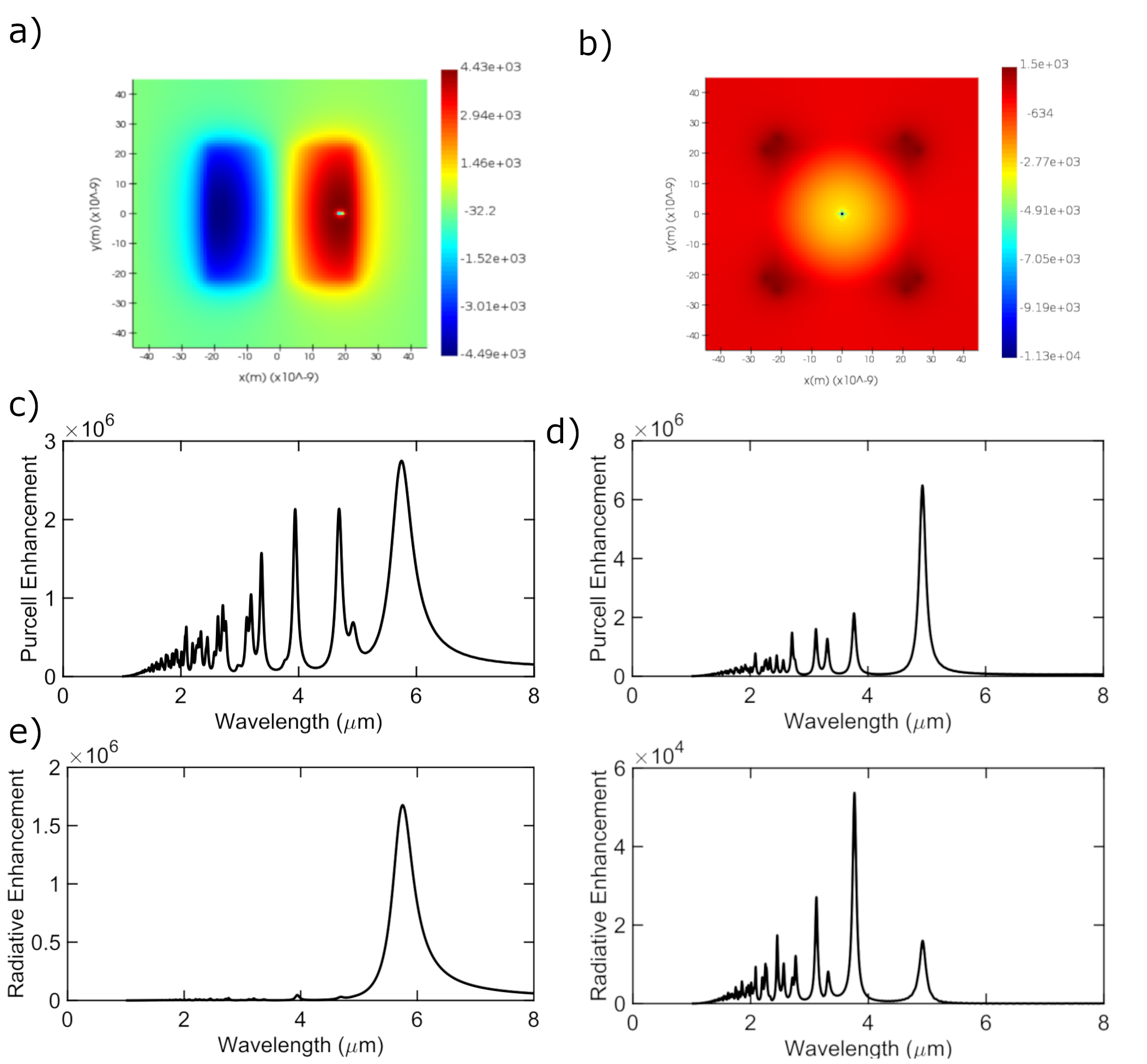}
    \caption{$E_Z$ field of the AGP at the center of the heterostructure cavity with respect to the z-axis for the case when the perpendicular dipole is placed both a) on resonance at $x$ = 18 nm and b) off resonance at $x$ = 0 nm. A large Purcell enhancement is seen for both c) off resonance and d) on resonance with both having a similar magnitude. However, we see that the radiative enhancement is much larger e) on resonance while radiative enhancement is much lower f) off resonance as the monopole excited by the dipole at $x$ = 0 nm does not couple strongly to the far field while the dipole resonances excited by the dipole at $x$ = 18 nm couples efficiently to the far field. This shows that the efficiency of the Purcell enhancement is highly dependent on the location of the emitter relative to the electric field distribution of the AGP.}
    \label{fig:PerpOnOff}
\end{figure*}

For both the perpendicular and parallel cases, we see several peaks in the Purcell factor corresponding to the allowed AGP eigenmodes in the cavity as depicted in Fig.~\ref{fig:ParallelOnOff} and Fig.~\ref{fig:PerpOnOff}. For the perpendicular orientation, a dipole emitter located at the geometric center of the hBN/WS$_2$/hBN heterostructure experiences a Purcell enhancement of several million. However, this monopole eigenmodes does not efficiently couple to the far field. As a result, we see radiative enhancement of less than 1 at the wavelength corresponding to this monopole. Suppose we instead shift the SPE's location in the x direction by $1/e$ of the side length of the nanocube. In that case, it is possible to excite a dipole $E_z$ field inside the cavity that couples more strongly to the far field resulting in radiative enhancement. The remaining peaks in the Purcell enhancement spectrum correspond to higher order eigenmodes with weak far-field coupling.
\par
For the parallel orientation, we are interested in the $E_x$ field component rather than the $E_z$ field component. Our FDTD calculations show the $E_x$ field to be a monopole localized under the nanocubes. Thus, we observe a relatively small electric field present when offset from the center of the nanocube and a corresponding sharp drop in the radiative enhancement when compared to an electric dipole source located under the center of the nanocube.

\subsection*{Atom-plasmonic cavity system}
The electric dipole emitter in the FDTD simulations models the spontaneous emission of an atom from the excited state $\left|e\right>$ to its ground state $\left|g\right>$. Since the electric dipole emitter is only a source of electromagnetic radiation, the FDTD simulations are valid in the weak coupling (Purcell) regime where the re-absorption and re-emission cycles are neglected (see e.g. Refs.~\onlinecite{Tafur2011,Tafur2012,Tafur_PhDthesis2011} for the detailed description of Rabi oscillations during the spontaneous emission process). 

Both the enhancement and the suppression of the spontaneous emission of a quantum emitter can be derived from Fermi's golden rule,\cite{Meystre2007}
\be
\Gamma=\frac{2\pi}{\hbar^2}\left|\left<f\left|\cH_{\rm int}\right|i\right>\right|^2 D(\omega)
=2\pi g^2(\br,\omega)D(\omega),
\ee
where $\cH_{\rm int}=-\hbd\cdot\hbE_{mn}$ is the perturbing interaction Hamiltonian, $\left|i\right>$ and $\left|f\right>$ are the initial and final states, respectively, and $g=\left<\cH_{\rm int}\right>/\hbar$ is the coupling frequency between the quantum emitter's dipole moment $\hbd=-e\hbr$ and the electric field $\hbE_{mn}$ of the AGP eigenmodes $(m,n)$.
The dipole moment operator between the ground state $\left|g\right>$ and excited state $\left|e\right>$ of the quantum emitter is given by
\be
{\bf{\hat d}} = \left| g \right\rangle {{\bf{d}}_{ge}}\left\langle e \right| + \left| e \right\rangle {\bf{d}}_{ge}^*\left\langle g \right|,
\ee
where 
${{\bf{d}}_{ge}} = \left\langle g \right|{\bf{\hat d}}\left| e \right\rangle 
=-e\left\langle g \right|\hbr\left| e \right\rangle$.
The electric field operator is 
\be
\hbE_{mn}(\br)=\bE_{mn}a_{mn}+\bE_{mn}^* a_{mn}\D,
\ee
where $a_{mn}$ and $a_{mn}\D$ are the annihilation and creation operators of the quantized AGPs in the eigenmodes $(m,n)$, respectively. The electric field components per single quantized AGP are given by\cite{Craft2023}
\begin{align}
    E_{mn,z}(\br) &= \sqrt{\frac{\hbar\omega_{mn}}{\varepsilon_0\varepsilon_r V_{mn}}} \sin\left(k_{mn,x} x\right)\sin\left(k_{mn,y} y\right)\cosh\left(\zeta_{mn,z} z\right) \nn\\
    E_{mn,x}(\br) &= \sqrt{\frac{\hbar\omega_{mn}}{\varepsilon_0\varepsilon_r V_{mn}}}\cos\left(k_{mn,x} x\right)\sin\left(k_{mn,y} y\right)\sinh\left(\zeta_{mn,z} z\right), \nn\\
    E_{mn,y}(\br) &= \sqrt{\frac{\hbar\omega_{mn}}{\varepsilon_0\varepsilon_r V_{mn}}}\sin\left(k_{mn,x} x\right)\cos\left(k_{mn,y} y\right)\sinh\left(\zeta_{mn,z} z\right),
\end{align}
where the eigenfrequencies of the AGPs are $\omega_{mn}=2\pi f_{mn}$, $V_{mn}$ is the optical mode volume, and $\varepsilon_r$ is the relative permittivity of the dielectric material between the Ag nanocube and graphene.
The density of states (DOS) $D(\omega)$ represents the number of states per unit energy that the quantum emitter can decay into at the frequency $\omega$. In vacuum the DOS in three dimensions is
\be
D_0(\omega)=\frac{\omega^2 V}{\pi^2 c^3},
\ee
where $c$ is the speed of light in vacuum.
Consequently, the spontaneous emission rate in vacuum is\cite{Meystre2007}
\be
\Gamma_0=\frac{\omega^3 d_{ge}^2}{3\pi\varepsilon_0\hbar c^3}.
\ee
Purcell realized that if the DOS is modified by a cavity, then the spontaneous emission rate can be varied.\cite{Purcell1946}
In the case of the AGP cavity the DOS is
\be
D_{\rm AGP}(\omega)=\frac{2}{\pi}\sum_{m,n}\frac{\gamma_{mn}}{4\left(\omega-\omega_{mn}\right)^2+\gamma_{mn}^2},
\ee
where $\gamma_{mn}$ are the linewidths of the Lorentzian peaks of the eigenmodes $(m,n)$. Thus, the spontaneous emission rate into AGP modes is
\be
\Gamma(\br,\omega) = 4g^2(\br,\omega)\sum_{m,n}\frac{\gamma_{mn}}{4\left(\omega-\omega_{mn}\right)^2+\gamma_{mn}^2}.
\ee
After choosing a particular eigenmode $(m_0,n_0)$ by setting $\omega\approx\omega_{m_0n_0}$, one obtains the enhancement/suppression factor
\begin{align}
F(\br,\omega) &=\frac{\Gamma(\br,\omega\approx\omega_{m_0n_0})}{\Gamma_0} \nn\\
&\approx
\frac{3Q_{m_0n_0}\left(\lambda_{m_0n_0}/n\right)^3}{4\pi^2 V_{m_0n_0}}\frac{\gamma_{m_0n_0}^2}{4\left(\omega-\omega_{m_0n_0}\right)^2+\gamma_{m_0n_0}^2} \nn\\
&\times\left(\frac{\bd\cdot\bE}{dE}\right)^2\frac{\left|\bE(\br)\right|^2}{\left|\bE_{\rm max}\right|^2},
\label{eq:Purcell}
\end{align}
where $Q_{m_0n_0}=\omega_{m_0n_0}/\gamma_{mn}$ is defined as the quality factor of the eigenmode $(m_0,n_0)$.
The prefactor of the above expression is the Purcell factor
\be
\calF\equiv F_{\rm max}=\frac{3Q_{m_0n_0}\left(\lambda_{m_0n_0}/n\right)^3}{4\pi^2 V_{m_0n_0}},
\label{eq:Purcell_max}
\ee
which is achieved for exact resonance $\omega=\omega_{m_0n_0}$, parallel or anti-parallel alignments of $\bd$ and $\bE$, i.e. $\bd\cdot\bE=\pm dE$, and maximum electric field $\bE(\br)=\bE_{\rm max}$.

\subsection*{Weak and strong-coupling regimes}
Eq.~\eqref{eq:Purcell} derived above is valid for the weak-coupling regime, in which $g < \kappa,\Gamma_0$. In the strong-coupling regime, the emitted photon goes through several reabsorption and reemission cycles, resulting in damped Rabi oscillations with a rate $g$. A quantum treatment gives the equation 
\be
\Gamma = \Gamma_0 + \frac{4g^2}{\kappa}
\ee
which we can rewrite in terms of the Purcell factor $\calF =\frac{\Gamma}{\Gamma_0}$ as
\be
\calF = 1 + \frac{4g^2}{\kappa\Gamma_0}
\ee
We can find the value of $\kappa$ for a given eigenmode from its frequency and quality factor as 
\be
\kappa=\frac{\omega_{m_0 n_0}}{Q_{m_0 n_0}}
\ee
We can then solve for $g/\kappa$ as 
\be
\frac{g}{\kappa}=\sqrt{\frac{(\calF-1) \Gamma_0}{4 \kappa}}=\sqrt{\frac{(\calF-1) \Gamma_0}{4} \frac{Q_{m_0 n_0}}{\omega_{m_0 n_0}}},
\label{eq:weak_strong_coupling}
\ee
which is a useful reformulation of Eq.~\eqref{eq:Purcell} as we can easily distinguish between the strong coupling regime $(g/\kappa > 1)$ and the weak coupling regime $(g/\kappa < 1)$.
For the IR cavity with a resonant frequency of $\omega_{m_0 n_0}=3.28\times 10^{14}$ rad/s, corresponding to a frequency of $f_{m_0 n_0}=\omega_{m_0 n_0}/2\pi=52.1$ THz and a wavelength of $\lambda_{m_0 n_0}=5.75$ $\mu$m, we have $\kappa = 3.9 \times10^{12}$ s$^{-1}$ and $\calF=2.75\times10^6$, and therefore
\be
\frac{g}{\kappa}=4.2 \times 10^{-4}\sqrt{\Gamma_0}.
\ee
Similarly, for the telecommunications wavelength cavity we have a resonant frequency of $\omega_{m_0 n_0}=1.22\times 10^{15}$ rad/s, corresponding to a frequency of $f_{m_0 n_0}=\omega_{m_0 n_0}/2\pi=193$ THz and a wavelength of $\lambda_{m_0 n_0}=1.55$ $\mu$m, we have $\kappa = 1.26 \times 10^{13}$ s$^{-1}$ and $\calF=1.76\times10^4$, and therefore
\be
\frac{g}{\kappa}=1.9 \times 10^{-5}\sqrt{\Gamma_0}.
\ee
Thus, we see that the strong coupling regime can be reached as long as the vacuum decay rate of the transition $\Gamma_0\gtrsim 5.7 \times 10^6$ s$^{-1}$ for $\lambda_{m_0 n_0}=5.75$ $\mu$m, and $\Gamma_0\gtrsim 2.3 \times 10^{9}$ s$^{-1}$ for $\lambda_{m_0 n_0}=1.55$ $\mu$m.

\subsection*{Electrostatic tuning of AGP frequency}
\par
It has been reported that the frequency of these AGPs can be tuned by means of a gate voltage applied to the graphene sheet.\cite{Craft2023} This gate voltage varies the Fermi energy of the graphene sheet and in turn the frequency of the AGP. We vary the Fermi energy in the FDTD simulation and change the corresponding scattering rate according to
\begin{equation}
  \tau^{-1}=\frac{e{v_F}^2}{\mu E_F}
\end{equation}
where $v_F = 10^6$ m/s is the Fermi velocity of graphene and $\mu$ is the mobility of the graphene sheet. This equation can also be rewritten in the standard form $\mu=e v_F^2 \tau / E_F$.
We vary the Fermi energy from $E_F = 0.8$ eV to  $1.2$ eV, while the thickness $h$ of the hBN/WS$_2$/hBN heterostructure is fixed at 5 nm, and a single graphene layer is present. When varying $E_F$, the frequency of the AGP resonances closely follows the square root dependence of $f_{\rm AGP}$ on $E_F$ as shown in Fig.~\ref{fig:near_square_root} and predicted in Eq.~\eqref{eq:AGP_eigenfrequency}. This method of tuning has the distinct advantage of being able to be varied in real time. In contrast, the thickness of the hBN/WS$_2$/hBN heterostructure and the number of graphene layers are fixed after device fabrication. We see that the AGP can be tuned electrostatically by approximately 2 $\mu$m by varying $E_F$ from 0.6 eV to 1.2 eV. This electrostatic tunability can be used to switch the Purcell enhancement on and off in real time by means of varying the applied gate voltage. By tuning between $E_F=0.6$ eV to $E_F=1$ eV we are able to achieve an on-off ratio of 19.7 dB in the radiative enhancement of the device.

\subsection*{Geometric tuning of AGP frequency}
\par
\begin{figure*}[htb]
    \centering
    \includegraphics[width=\textwidth]{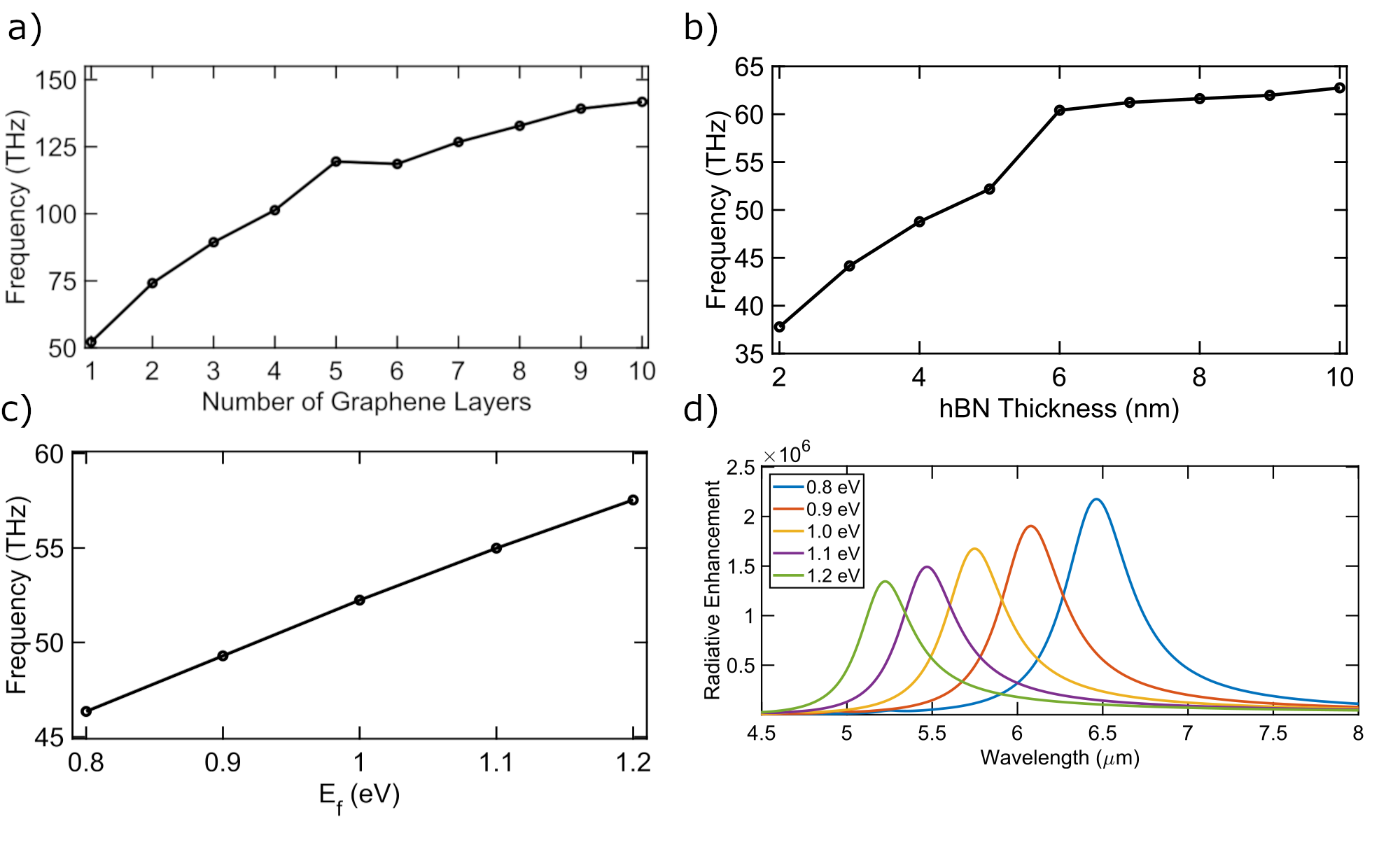}
    \caption{Near square root relationship between the frequency of the excited AGP in the cavity as a function of a) number of graphene layers intercalated with FeCl3 ($C_S$), b) hBN layer thickness ($h$), and c) graphene Fermi energy ($E_F$) for the perpendicular dipole orientation. The radiative enhancement factor as a function of wavelength is shown for d) multiple values of $E_F$ as well to clearly demonstrate the electrostatic tunability of the AGPs. When not the variable of interest, we set $C_S$ = 1, $E_F$ = 1.0 eV, and $h$ = 5 nm.}
    \label{fig:near_square_root}
\end{figure*}

The AGP frequency is predicted to show a square root dependence on the thickness of the hBN/WS$_2$/hBN heterostructure according to Eq.~\eqref{eq:AGP_eigenfrequency}. We calculate the AGP frequency using Lumerical FDTD and vary the thickness of the hBN/WS$_2$/hBN heterostructure from 1 nm to 10 nm. We fix the Fermi energy at $E_F=1$ eV and use a single graphene layer. We see in Fig.~\ref{fig:near_square_root} that the rate of increase in frequency of the AGP resonances decreases rapidly from the expected square root dependence at larger thickness of the hBN/WS$_2$/hBN heterostructure. However, as this thickness decreases, we approach the expected square root dependence.
\par
In addition, we propose using multi-layered graphene to further increase the frequency of the AGPs. This is accomplished using multiple layers of graphene intercalated with FeCl$_3$. With this method, the distance between the graphene layers is increased resulting in very weak interactions between the graphene layers. \cite{https://doi.org/10.1002/adfm.201000641} As a result, we expect the total conductivity of the multi-layer graphene intercalated with FeCl$_3$ to scale with the number of graphene layers. The AGP frequency has been shown to scale with the square root of the total conductivity of graphene. \cite{gonccalves2021quantum} As a result, we expect the AGP frequencies to scale with the square root of the number of graphene layers $C_S$. We vary $C_S$ in the Lumerical FDTD simulation and observe the change in AGP frequency and Purcell enhancement factor $\calF$. We see that for the dipole AGP eigenmode, the relationship between AGP frequency and $C_S$ scales with a near square root relationship.

\subsection*{Effects of tuning methods on enhancement factor}
\par
\begin{figure*}[htb]
    \includegraphics[width=\textwidth]{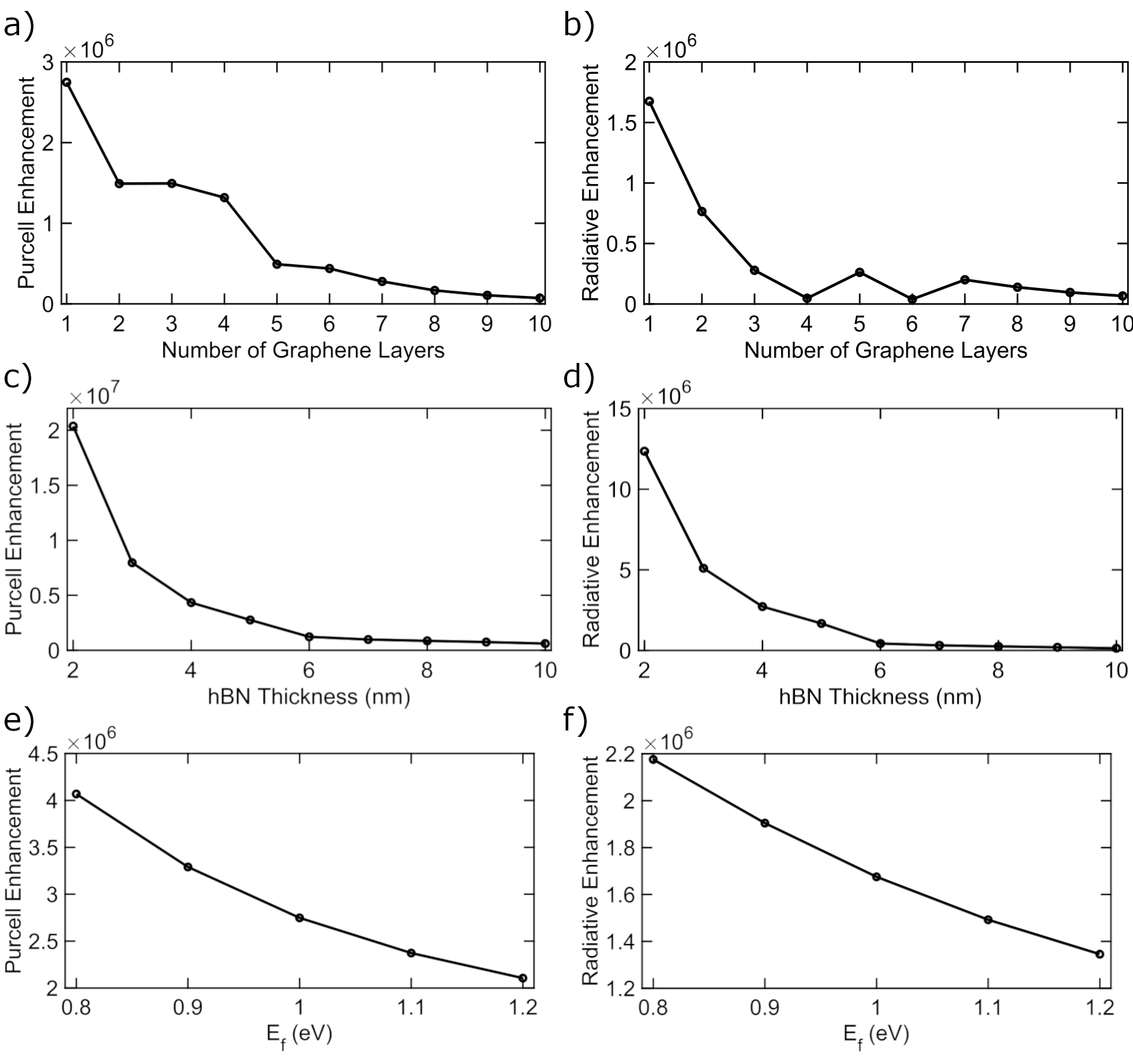}
    \caption{Relationship between the number of intercalated graphene layers (Cs) and the a) peak Purcell enhancement and b) peak radiative enhancement.  Relationship between the hBN layer thickness (h) and the c) peak Purcell enhancement and d) peak radiative enhancement. Relationship between the graphene Fermi energy and the c) peak Purcell enhancement and d) peak radiative enhancement. When not the variable of interest, we set $C_S$ = 1, $E_F$ = 1.0 eV, and $h$ = 5 nm.}
    \label{fig:Enhancement}
\end{figure*}

While all three of the presented methods for tuning the AGP frequency scale with a near square-root dependence, they differ in how they affect the magnitude of the enhancement factor. The radiative transition rates and Purcell enhancement factors as a function of $E_F$, $h$, and $C_S$ for the perpendicular dipole orientation are shown in Fig.~\ref{fig:Enhancement}. In general, we see that increasing the frequency of the AGP decreases the enhancement factor experienced by the dipole source.  Although increasing the value of $C_S$ and $h$ initially result in a large decrease in enhancement, we see that both reach a point at which further increases have diminishing effects on enhancement. This is promising as it allows us to greatly increase the frequency of the AGPs while maintaining a high enhancement rate, which, as we see later, allows us to reach telecommunications wavelengths while maintaining a large Purcell enhancement.

\subsection*{Example of frequency tuning to 1550 nm}
\begin{figure}[htb]
    \includegraphics[width=3.5 in]{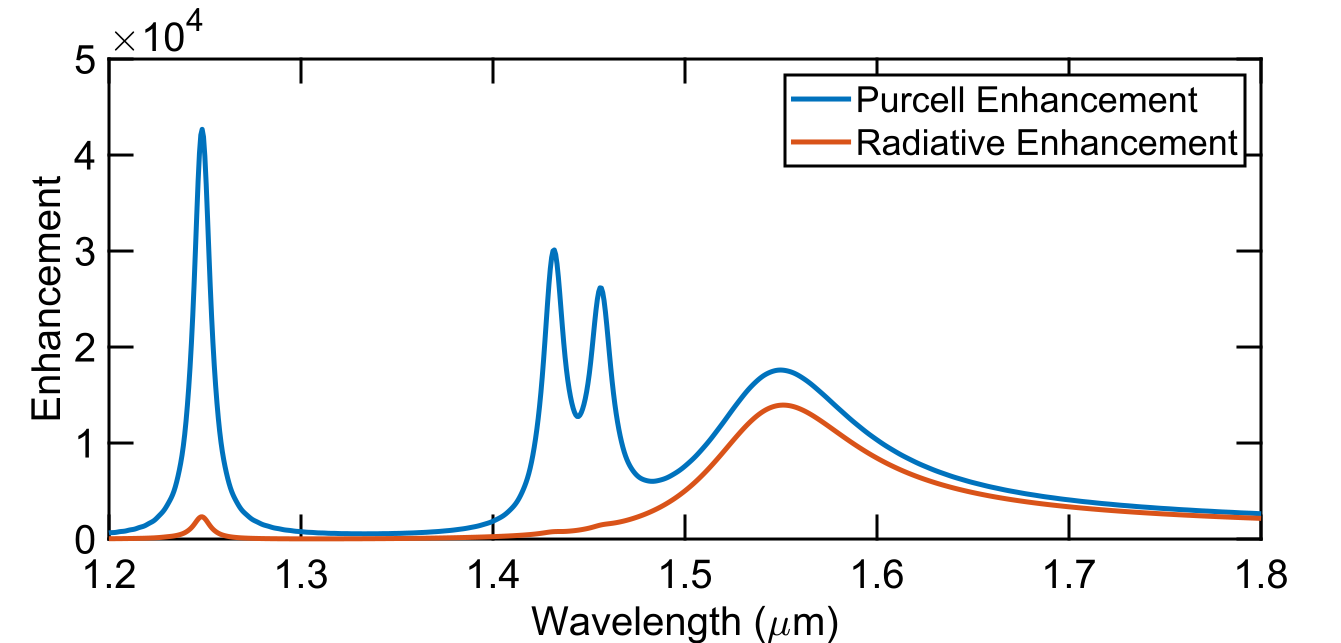}
    \caption{Demonstration of  tuning of electric dipole emission for $C_S$ = 10, $h$ = 10 nm, and $E_F$ = 1.2 eV to enable operation in the C band at 1550 nm  with a Purcell enhancement of $1.76\times 10^4$.}
    \label{fig:1550}
\end{figure}

To demonstrate the tuning capability of the proposed heterostructure, we tune the AGP resonance to 1550 nm as shown in Fig.~\ref{fig:1550}. For the hBN/WS$_2$/hBN heterostructure, we use a thickness of 10 nm, $C_S$ = 10 graphene layers, and $E_F = 1.2$ eV. As noted previously, achieving this high frequency results in a reduction in the enhancement factors. Using the aforementioned parameters, we achieved an AGP wavelength of 1550 nm with a Purcell enhancement factor of $\calF=1.76\times10^4$ and a quantum efficiency of $QE=0.79$ for a total radiative enhancement factor of $\xi_{\rm ON}=\calF QE=1.39\times10^4$. We can shift the AGP frequency, and therefore the peak in enhancement, to higher wavelengths by decreasing the Fermi energy. Doing so allows for the enhancement of the 1550 nm electric dipole to be effectively switched on and off. By setting the Fermi energy to $E_F = 0.75$ eV, we can reduce the radiative enhancement factor to $\xi_{\rm OFF}=40.36$. This means that the radiative enhancement factor can be varied by a factor of 
\be
\frac{\xi_{\rm ON}}{\xi_{\rm OFF}}=3.44\times 10^2
\ee
between the on and off modes of operation, which is equivalent to $10\log_{10}(\xi_{\rm ON}/\xi_{\rm OFF})=25.4$ dB.

\subsection*{Effects of graphene mobility on QE}
\begin{figure}[htb]
    \centering
    \includegraphics[width=0.5\textwidth]{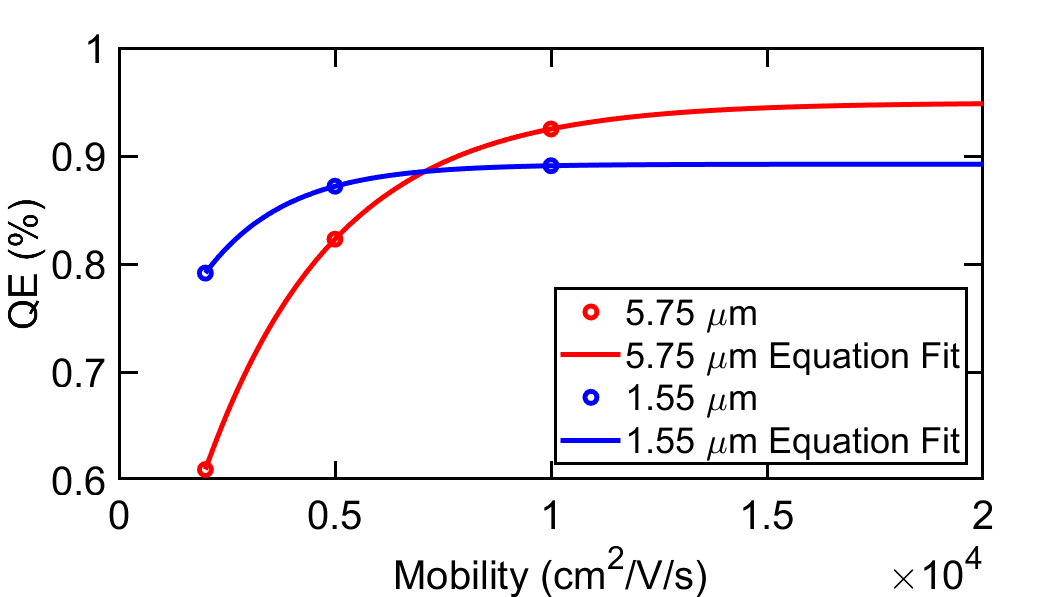}
    \caption{Quantum efficiency of the cavity for a resonance centered at 1.55 $\mu m$ and 5.75 $\mu m$ as a function of the graphene mobility along with best fit equations.}
    \label{fig:QEfit}
\end{figure}

The previous results in this study have assumed a graphene mobility of $\mu=2000$ cm$^2$/Vs, which is a relatively conservative value for modern fabrication methods. Graphene mobilities of up to $\mu=125,000$ cm$^2$/Vs have been reported in hBN encapsulated graphene \cite{10.1063/1.3665405}, and up to $\mu=10,000$ cm$^2$/Vs using scalable fabrication methods\cite{doi:10.1021/acsami.3c06120}. We expect that using a higher mobility increases the QE of the device. In Fig. ~\ref{fig:QE}, we see that increasing the mobility to $\mu=10,000$ cm$^2$/Vs results in a $QE=92.4\%$ for $\lambda=5.75$ $\mu$m and $QE=89\%$ for $\lambda=1.55$ $\mu$m. This increase in quantum efficiency $QE$ is vital for creating high quality single photon and entangled photon sources, as it directly determines the value of $\beta_{ext}$. In summary, this shows that high quantum efficiency $QE$ is achievable in AGP resonances if high mobility graphene is used.  We expect that the QE must saturate at some point, as it can never excede 100\%. Thus, we fit the data to a function of the form $1-e^{-x}$. This analysis suggests that the quantum efficiency $QE$ of the device optimized for $\lambda=5.75$ $\mu$m saturates at 95\% and for the $\lambda=1.55$ $\mu$m device $QE$ saturates at 89\% as shown in Fig. ~\ref{fig:QEfit}.

\subsection*{Extension to electric and magnetic multipolar and nonlinear optical transitions}
The Lumerical FDTD software does not contain methods for simulating multipolar transitions, two-plasmon spontaneous emission, or spin-flip transitions. However, it has been shown that the magnitude of fluorescence enhancement for the electric dipole transition can be used to calculate the enhancement of these other emission processes \cite{Rivera&Joannopoulos&Soljacic2016}. The rate enhancement of each transition is calculated from a characteristic value $\eta$ according to the equations shown in Table~\ref{tab:enhancement_factors}. We see that the equation for the enhancement of the electric dipole ($E1$) transition is equal to $\eta ^3$, and we can thus extract the value of $\eta$ as the cube root of our Purcell enhancement factor, i.e.
\be
\eta=\calF^{\frac{1}{3}}. 
\ee
Using this identity, we can  calculate the enhancement rate of the other transition types based on the characteristic equation of that transition using our value of $\eta$. Examples are given in Table~\ref{tab:enhancement_factors} for two different wavelengths. The enhancement factors for the multipolar and higher-order transitions are much larger than for the electric dipole transition, enabling the measurement of transitions that are not typically visible experimentally.

\begin{table}[htb]
\begin{tabular}{|p{2cm}|p{2cm}|p{2cm}|p{2cm}|}
 \hline
 \multicolumn{4}{|c|}{Purcell Enhancement Factors for Various Transition Types} \\
 \hline
 Optical Transition & Enhancement Factor Formula & Enhancement Factor Value $\lambda$ = 5.75 $\mu m$ & Enhancement Factor Value $\lambda$ = 1550 $nm$\\
 \hline
 $E1$   & $\calF=\eta^3$ & $2.75\times 10^{6}$ & $1.76\times10^{4}$\\
 \hline
 $E2$ &   $\eta^{3+2(n-1)}$  & $5.4\times 10^{10}$ & $1.19\times10^{7}$\\
 \hline
 $E3$ & $\eta^{3+2(n-1)}$ & $1.06\times 10^{15}$ & $8.06\times10^{9}$\\
 \hline
 $2PSE$ & $\eta^6$ & $7.56\times 10^{12}$ & $3.1\times10^{8}$\\
 \hline
 $M1$ & $\eta^{1+2(n-1)}$ & $140.1$ & $26.0$\\
 \hline
 $M2$ & $\eta^{1+2(n-1)}$  & $2.75\times 10^{6}$ & $1.76\times10^{4}$\\
 \hline
 $M3$ & $\eta^{1+2(n-1)}$  & $5.4\times 10^{10}$ & $1.19\times10^{7}$\\
 \hline
 \end{tabular}
 \caption{Enhancement rates and rate equations for several transition types including multipolar electric ($E1$: electric dipole, $E2$: electric quadrupole, $E3$: electric octupole), magnetic ($M1$: magnetic dipole, $M2$: magnetic quadrupole, $M3$: magnetic octupole), and 2-photon spontaneous emission ($2PSE$) based on electric dipole ($E1$) matrix elements. The order of the multipolar transition in the enhancement factor is identified by $n$.}
 \label{tab:enhancement_factors}
 \end{table}
 
\subsection*{Dark and bright modes in the AGP cavity}
Of the several modes supported by the AGP cavity, only one of these modes is bright while the others are dark, i.e. the dipole AGP mode is the sole bright mode. This mode couples strongly to the far-field with quantum efficiencies $QE$ of up to 90\%. In contrast, the next higher order modes are dark modes. These dark modes only couple a small fraction of emitted energy to the far-field. The remaining energy emitted by dark modes is primarily coupled into decaying plasmon modes. This energy decays into vibrational modes that dissipate as heat into the graphene layer. It has been shown that when multiple plasmon cavities are in close proximity, it is possible to have couplings between plasmons that result in the shifting of the plasmon frequencies, or even splitting into two hybridized modes.\cite{doi:10.1021/nl049681c} While we do not observe this splitting for the bright dipole mode, the higher order dark plasmon modes do exhibit this splitting. The frequency difference of these peaks is constant with respect to the vacuum decay rate of the emitter, suggesting that this is indeed due to plasmon hybridization and not strong coupling between the SPE and the plasmon cavity.

\subsection*{Potential applications}
\subsubsection*{Telecommunication bands used in optical fibers}
Rare-earth atoms have been used for classical communication in the telecommunication bands shown in Table~\ref{tab:fiber_bands}.

\begin{table}[htb]
\centering
\begin{tabular}{ |c|c|c| }
 \hline
 \textbf{Band} & \textbf{Wavelength [nm]} & \textbf{Description} \\
 \hline
 O & 1260–1360 & Original band \\
 \hline
 E & 1360–1460 & Extended band \\
 \hline
 S & 1460–1530 & Short wavelength band \\
 \hline
 C & 1530–1565 & Conventional band \\
 \hline
 L & 1565–1625 & Long wavelength band \\
 \hline
 U & 1625–1675 & Ultralong wavelength band \\
 \hline
\end{tabular}
\caption{Standard bands for optical fiber communications}
\label{tab:fiber_bands}
\end{table}

\begin{figure}[htb]
    \centering
    \includegraphics[width=0.5\textwidth]{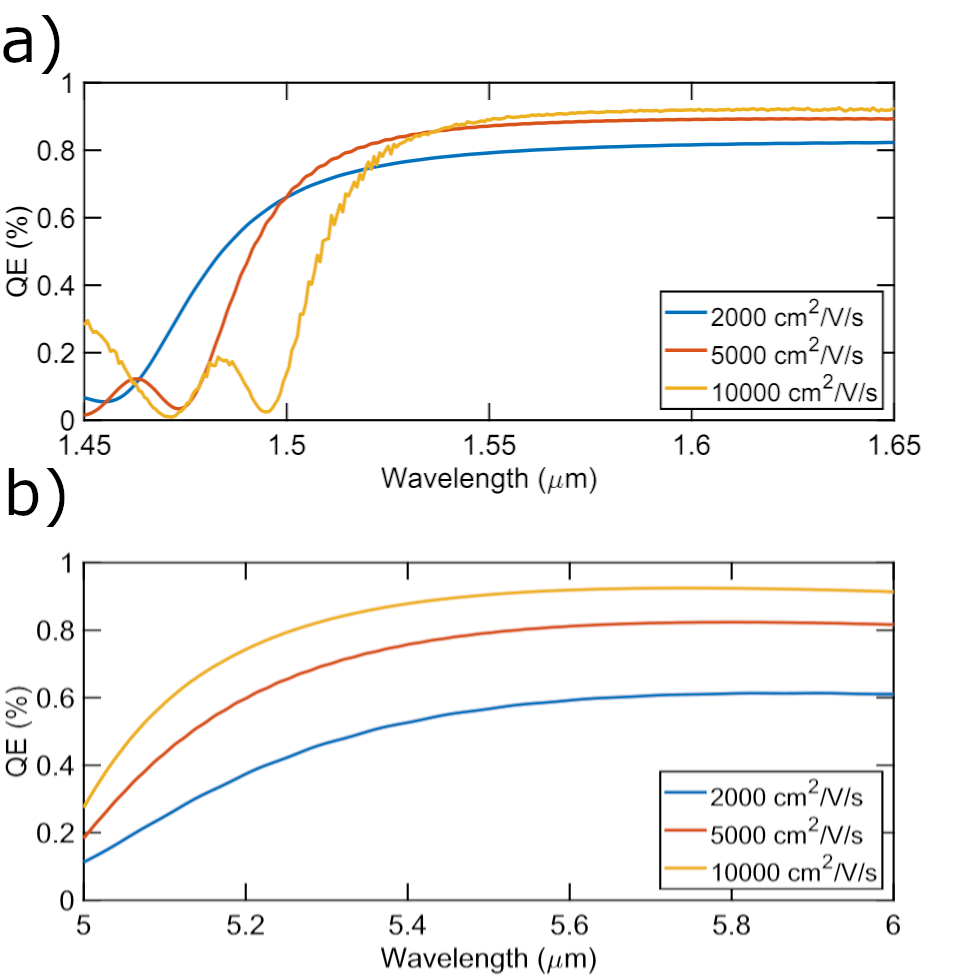}
    \caption{Quantum efficiency of the cavity for a resonance centered at (a) 1.55 $\mu m$ and (b) 5.75 $\mu m$.}
    \label{fig:QE}
\end{figure}

\subsubsection*{Single-photon source based on electric dipole transitions in rare-earth-doped TMDs}
It has been shown that rare-earth doped 2D TMDs are candidates for photoluminescence sources in the NIR region, including at 1550 nm. \cite{https://doi.org/10.1002/adma.201601833, C8NR01139G,doi:10.1021/acs.jpcc.2c07113} It is possible to dope WS$_2$ with Er, where the Er ions are substituted into the WS$_2$ structure by replacing W cations. This Er:WS$_2$ material exhibits a fluorescence transition at approximately 1550 nm \cite{https://doi.org/10.1002/adma.201601833,doi:10.1021/acs.jpcc.2c07113,khan2021ab}. We propose to use our hBN/WS$_2$/hBN heterostructure with this Er:WS$_2$ material as shown in Fig.~\ref{fig:schematic_Er_doping}. Similar results can be expected for hBN/MX$_2$/hBN heterostructure, where M=Mo, W is the transition metal atom, and X=S, Se is the chalcogen atom. We then expect this fluorescence transition to have a greatly enhanced transition rate. 

First we show that Er:WS$_2$ indeed shows a transition at 1.5 microns by using first principle methods. All numerical calculations are performed using density functional theory (DFT) with the hybrid generalized gradient approximation (GGA) in the form of the HSE functional, as implemented in the Synopsis Atomistix Toolkit (ATK) 2022.\cite{QuantumATK} In hybrid GGA methods, a portion of the Hartree–Fock exchange is combined with the GGA exchange-correlation functional. This mixing fraction is typically denoted by "$\alpha$", with values ranging from 0 (pure GGA) to 1 (pure Hartree–Fock). In this work, we employ the standard HSE06 hybrid functional parameters ($\alpha=0.25$). The inclusion of Hartree–Fock exchange in hybrid GGA calculations leads to an improved description of electronic properties such as band gaps, which are often underestimated in conventional GGA approaches. The system under consideration is modeled using a supercell of WS$_2$ comprising 63 W atoms, 128 S atoms, and one substitutional Er impurity (Er$_\mathrm{W}$). A vacuum layer of 25~\AA is included to screen interlayer interactions. Prior to the electronic structure calculations, the atomic positions in the supercell are fully relaxed until the maximum force on each atom is below 0.01~eV/\AA.

\begin{figure*}[htb]
    \centering
    \includegraphics[width=\textwidth]{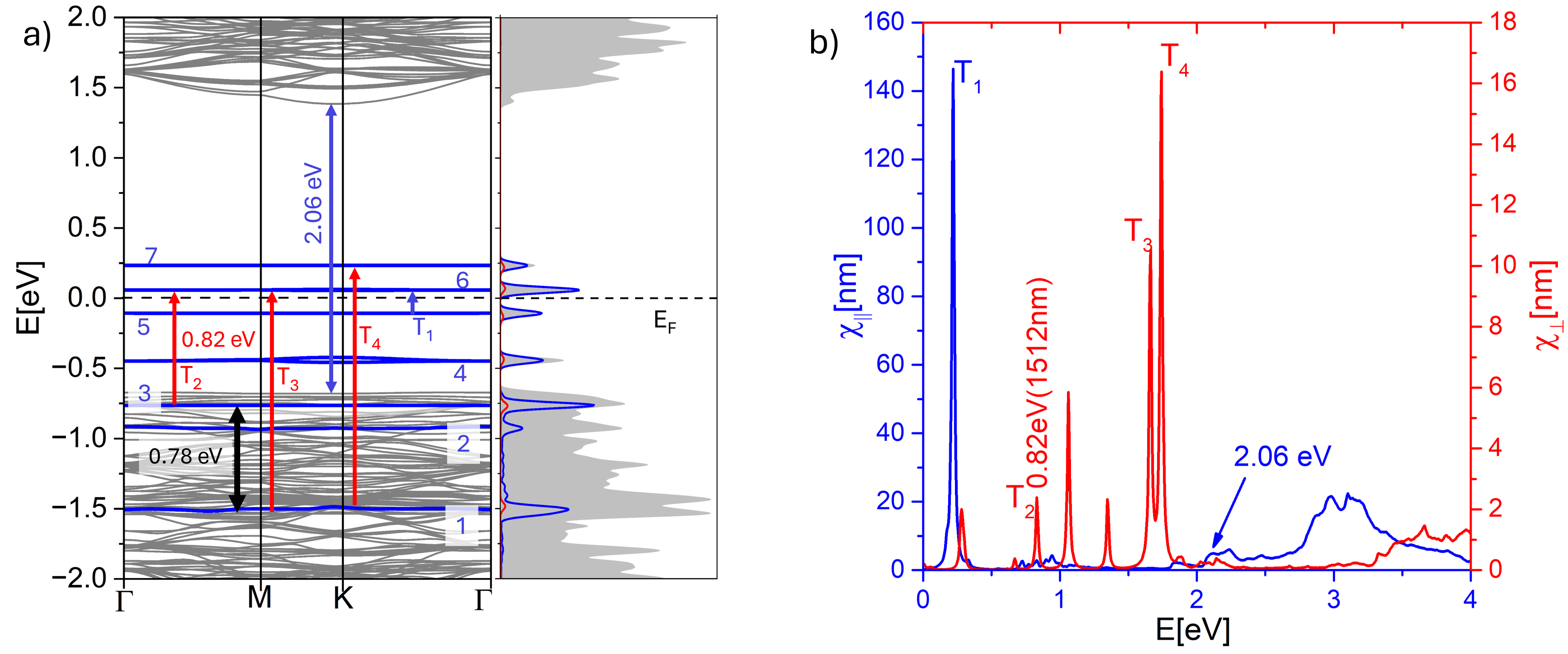}
    \caption{Band structure, density of states (DOS), and optical spectrum of Er-doped single-layer WS$_2$. The blue horizontal lines in the band structure and the blue peaks in the DOS represent the localized 4$f$ impurity states of Er. The red curves in the DOS indicate the finite contribution from the Er 5$d$ orbitals, while the shaded grey region denotes the total DOS. Colored vertical arrows mark representative optical transitions between localized defect states (LDS) observed in the susceptibility tensor, corresponding to in-plane (blue) and out-of-plane (red) polarization components.}  
    \label{BS_DOS_OS_ErAl}
\end{figure*}

The band structure and the density of states for the Er doped SL WS$_2$ are shown in the Fig.~\ref{BS_DOS_OS_ErAl}. The projected bandstructure and density of states on 4f orbitals are also shown Fig.~\ref{BS_DOS_OS_ErAl} (a) (blue color), appearing as dispersionless atomic like impurity states. Also, a finite contribution of 5d-orbitals (red color) can also be seen along with the f-orbitals in the density of states plot. Er ions have a characteristic signature emission in the 1.5 $\mu$m (0.82 eV) range and have been reported in various studies involving Er dopants in semiconductors and insulators.\cite{khan2021ab, khan2025er}  Remarkably we are able to reproduce the characteristic radiative transition between different 4f orbitals of Er in the optical spectrum Fig.~\ref{BS_DOS_OS_ErAl} (b) as T$_1$ at 0.82 eV ($\sim$ 1.52 microns), The appearance of characteristic emission of Er in different crystal environments such as SL WS$_2$ shows the minimum coupling of 4f orbitals of Er with the host material such as SL WS$_2$.

Radiative transitions among $f$-orbitals have been extensively investigated since the 1960s, as such processes are nominally forbidden by electric dipole selection rules. The Judd–Ofelt theory~\cite{Judd, Ofelt} provides a framework to describe these transitions by introducing odd-parity terms into the crystal-field Hamiltonian, which arise only in non-centrosymmetric crystal environments. Within the second-order perturbation framework, it has been shown that $f$-orbitals can hybridize with 5$d$-orbitals, i.e., $\langle f | \hat{H}_{\text{cry}} | d \rangle \neq 0$, only when $\hat{H}_{\text{cry}}$ contains odd-parity components—thus enabling otherwise forbidden radiative transitions. In the present case, the Er impurity in WS$_2$ occupies a site of D$_{3h}$ symmetry, which lacks inversion symmetry and therefore allows odd terms in the crystal-field Hamiltonian. The corresponding crystal-field potential can be expressed as
\begin{equation}
\hat{H}_{\mathrm{CF}} = B_2^0 \hat{O}_2^0 + B_3^3 \hat{O}_3^3 + B_4^0 \hat{O}_4^0 + \dots
\end{equation}
The odd-parity component $B_3^3 \hat{O}_3^3$ arises exclusively from the non-centrosymmetric nature of the D$_{3h}$ point group and is responsible for mediating the coupling between the 4$f$ and 5$d$ orbitals of Er, as reflected in the blue and red features of the density of states shown in Fig.~\ref{BS_DOS_OS_ErAl}(a).

Using the Kohn-Sham states, we can calculate the spontaneous emission rate for the fluorescence transition at about 1550 nm.
The spontaneous emission rate of a photon can be derived using Fermi's golden rule, as shown in the Supplementary Information. We use the following relation to calculate the single-photon emission rate $1/\tau$
\begin{equation}
    \frac{1}{\tau}=\frac{4}{3\epsilon^{\perp}_r}\left(\frac{e^2}{4\pi\epsilon_0\hbar c}\right)\left(\frac{\omega^3_{3,6}}{c^2}\right)|\langle3|z|6\rangle|^2
\end{equation}
$\omega_{3,6}$ denotes the frequency of radiation corresponding to the transition from level $|6\rangle$ to level $|3\rangle$, as illustrated in Fig.~\ref{BS_DOS_OS_ErAl}(a), while $\langle3|z|6\rangle$ represents the electric dipole matrix element between the electronic states $|6\rangle$ and $|3\rangle$. Within the electric dipole approximation, the calculated spontaneous emission rate for this transition is found to be $1/\tau=2.6 \times 10^5$ s$^{-1}$. Inserting this value for $\Gamma_0=1/\tau$ in Eq.~\eqref{eq:weak_strong_coupling}, we obtain $g/\kappa=0.01$, which is clearly in the Purcell regime.

Based on our Purcell enhancement factor of $1.76\times10^4$ $\sim$ $1550$ nm, we can reduce this lifetime to $2.185\times 10^{-10}$ s. However, we must note that the quantum efficiency is also reduced by a factor of 0.788. This gives an effective radiative lifetime of $2.77\times 10^{-10}$ s. This large, tuneable radiative enhancement has the potential to be used to modulate the intensity of emitted light at 1550 nm, making it a promising candidate for use in fiber optic telecommunication systems. In particular, rare-earth atoms have typically very long lifetimes. Our heterostructure shows that these long lifetimes can be reduced by factors of the order of $10^4$.

\subsubsection*{Single-photon source based on weakly Purcell-enhanced magnetic dipole transitions in rare-earth-doped TMDs}
Magnetic dipole (M1) transitions in rare-earth atoms have typically very long lifetimes, making them very challenging to observe experimentally. Inspecting the tables of M1 transitions in rare-earth atoms in Ref.~\onlinecite{Dodson2012}, we can identify the following M1 transitions with vacuum rates $>1.0$ s$^{-1}$ and wavelengths $1400$ nm $<\lambda<1700$ nm that can be enhanced by means of AGPs. According to Table~\ref{tab:enhancement_factors}, we obtained an estimate of the Purcell enhancement factor of $\calF_{\rm M1}=37.5$. This results in relatively weakly Purcell-enhanced M1 emission rates, shown in Sec.~VII in SI. %Table~\ref{tab:M1_enhancement_factors}.

\subsubsection*{Single-photon source based on strongly Purcell-enhanced electric quadrupole transitions in rare-earth-doped TMDs}
Electric quadrupole (E2) transitions involve a change in angular momentum of two units, which cannot be accounted for by the photon's spin alone. To conserve total angular momentum, the emitted photon must carry orbital angular momentum (OAM) in addition to its intrinsic spin, i.e. $J=2$. This angular momentum arises from the spatial structure of the electromagnetic field, such as helical phase fronts in the emitted light. Photons with OAM are important because they enable access to higher-order transitions, allow for encoding information in new degrees of freedom,\cite{erhard2018twisted} and are essential in advanced applications like quantum communication\cite{yao2011orbital, PhysRevX.15.011024} and high-resolution spectroscopy. Understanding and utilizing this angular momentum opens new frontiers in light–matter interaction and photonic technologies.

Electric quadrupole (E2) transitions in rare-earth atoms have typically very long lifetimes, making them very challenging to observe experimentally. 

We found above that we can achieve a Purcell enhancement rate of $\calF=1.76\times10^4$ at 1550 nm for the electric dipole ($E1$) transition. From this, we find the quadrupole Purcell enhancement to be $1.19\times 10^{7}$.

Inspecting the tables of E2 transitions in rare-earth atoms in Ref.~\onlinecite{Dodson2012} (see Sec.~VII in SI), we can identify the following E2 transitions with vacuum rates of the orders of $10^{-5}$ s$^{-1}$ to $10^{-4}$ s$^{-1}$ and wavelengths $1400$ nm $<\lambda<1700$ nm that can be enhanced by means of AGPs. According to Table~\ref{tab:enhancement_factors}, we obtained an estimate of the Purcell enhancement factor of $\calF_{\rm E2}=1.19\times 10^{7}$. This results in giant Purcell-enhanced E2 emission rates shown in Sec.~VII in SI, ranging between $154$ s$^{-1}$ and $4573$ s$^{-1}$.

\subsubsection*{Entangled-photon source based on rare-earth atoms in TMDs}
2-photon electric dipole (2PSE) transitions can result in the emission of two entangled photons. These transitions are possible in rare-earth atoms in TMDs at wavelengths of interest to telecommunications \cite{PhysRevB.107.195420,PhysRevB.86.125102}. As discussed previously, we can predict the Purcell enhancement rate of these transitions based on our simulated Purcell enhancement rate of the dipole transition \cite{Rivera&Joannopoulos&Soljacic2016}. 
%We found previously that we can achieve a Purcell enhancement rate of $\calF=5.29\times10^4$ at 1550 nm for the electric dipole ($E1$) transition. 
For 2PSE transitions, it has been predicted that Erbium has a 2PSE transition with a vacuum decay rate of $2.943\times10^{-7}$ s$^{-1}$ at a wavelength of $\lambda=3.1$ $\mu$m. 
%Also, a formula for the 2PSE enhancement of this transition has been derived, and is based on the electric dipole Purcell enhancement in each of 3 perpendicular axes according to Ref.~\onlinecite{PhysRevB.107.195420}. That formula in the case of Er$^{3+}$ for two-photon emission from the excited state $^4$I$_{13/2}$ to the ground state $^4$I$_{15/2}$ is
%\begin{equation}
 % \calF_{\rm 2PSE}(\omega,\br)=\frac{\gamma(\omega,\textbf{r})}{\gamma_0(\omega)}=\sum_{i,j}A_{i,j}\calF_{i}(\omega,\textbf{r})\calF_{j}(\omega_0-\omega,\textbf{r})
%\end{equation}
%with
%\begin{equation}
 % A = 
%\begin{pmatrix}
%2/15 & 1/10 & 1/10\\
%1/10 & 2/15 & 1/10\\
%1/10 & 1/10 & 2/15
%\end{pmatrix}
%\end{equation}
%where $i,j=x,y,z$ are the set of three orthogonal axes, and $\calF_{i}(\omega,\textbf{r})$ is the Purcell enhancement along axis $i$ at frequency $\omega$ and position $\br$. 

Using the Purcell enhancement factor for localized AGPs, we obtain that the Purcell enhancement for a 2PSE emitter at $\lambda=3.1$ $\mu$m is of the order of $10^{12}$, two orders of magnitude larger than for the case of graphene only (see Ref.~\onlinecite{PhysRevB.107.195420}).

\subsection*{Entangled Photon Emission at telecommunications wavelengths}
The energy difference between the states $|6\rangle$ and $|3\rangle$ is 0.82 eV while the energy difference between $|3\rangle$ and $|1\rangle$ is 0.78 eV, as shown in Fig:~\ref{BS_DOS_OS_ErAl}. This 3 level system can be employed for a two-photon emission with intermediate energies i.e. $\hbar\omega+\hbar\omega^{\prime}=1.6$ eV, by considering $|3\rangle$ as the intermediate state. We use a second order time dependent perturbation theory to calculate the two-photon emission decay rate i.e. (see SI)
\begin{align}
\frac{1}{\tau^{(2)}} &= 128 \pi^5 \left( \frac{e^2}{\sqrt{\epsilon_{\parallel}\epsilon_{\perp}}\hbar c} \right)^2 
    \left( \frac{\omega_{63}^5}{c^4} \right) 
    \left| \langle 1 | x_{\parallel} | 3 \rangle \langle 3 | x_{\perp} | 6 \rangle \right|^2
\nn\\
&\times    \int_{0}^{2} \frac{x(2 - x)}{(1 - x)^2} \, dx
\end{align}
The matrix elements $\langle p|x_{\parallel(\perp)}|q\rangle$ were obtained from first-principles calculations. Using the values, the two-photon emission rate was calculated to be $9.127\times10^{-5}$~s$^{-1}$. By incorporating the Purcell enhancement factor $3.1\times10^8$ and accounting for a reduction in the quantum yield, the corresponding two-photon emission rate becomes $\frac{1}{\tau^{(2)}}=2.23\times 10^4$~s$^{-1}$, which corresponds to a lifetime  of $\tau^{(2)}=4.49\times10^{-5}$ s. This large, tunable radiative enhancement enables modulation of the emission intensity $\sim$ 1550~nm, positioning the system as a strong candidate for integration into fiber-optic telecommunication technologies.

\subsection*{Cooperativity and $\beta$-factor}

The total radiative enhancement factor $\xi$ accounts for both the Purcell enhancement factor $\calF$ and the quantum efficiency $QE$. The Purcell enhancement factor $\calF$ describes the enhancement of the radiative decay rate due to the presence of an optical cavity, i.e.
\begin{equation}
\mathcal{F} = \frac{\Gamma_{\text{rad}}}{\Gamma^{0}_{\text{rad}}}.
\end{equation}
Thus, the total radiative enhancement factor is given by
\begin{equation}
\xi = \mathcal{F} QE = \frac{\Gamma_{\text{rad}}}{\Gamma^{0}_{\text{rad}}} \cdot \frac{\Gamma_{\text{rad}}}{\Gamma_{\text{rad}} + \Gamma_{\text{loss}}}
\end{equation}
This factor represents the combined effect of the Purcell enhancement and the quantum efficiency in determining the overall radiative enhancement of the quantum emitter.

The cooperativity and $\beta$-factor are important quantities that provide information about the quality of the extracted quantum light.
On resonance and for optimal dipole alignment, we define the cavity–emitter cooperativity as
$C \equiv 4g^{2}/(\kappa\,\Gamma_{0})$, where $g$ is the vacuum Rabi frequency, $\kappa$ the cavity energy-decay rate, and $\Gamma_{0}$ the free-space spontaneous-emission rate. The cavity-enhanced decay rate is then $\Gamma=\Gamma_{0}(1+C)$, i.e.\ the Purcell enhancement $\mathcal{F}\equiv \Gamma/\Gamma_{0}=1+C$. The fraction of emission funneled into the bright cavity mode (the $\beta$-factor) is
\be
\beta_{\mathrm{cav}}=\frac{C}{\mathcal{F}}=\frac{C}{1+C}=\frac{\mathcal{F}-1}{\mathcal{F}}.
\ee
Including dissipative loss via the quantum efficiency $QE\equiv \Gamma_{\mathrm{rad}}/(\Gamma_{\mathrm{rad}}+\Gamma_{\mathrm{loss}})$, the far-field (device-relevant) branching ratio is
\be
\beta_{\mathrm{ext}}=\beta_{\mathrm{cav}}\times QE\simeq QE\quad(C\gg 1).
\ee
Thus, approaching deterministic single-photon emission corresponds to $\beta_{\mathrm{ext}}\to 1$; for symmetric two-photon emission, the pair-in-mode probability scales as $\beta_{\mathrm{pair}}\simeq \beta_{\mathrm{ext}}^{2}$. Using high-mobility graphene, for the mid-IR cavity design, we have $\beta_{\mathrm{ext}}=QE=0.95$ and for the telecommunications cavity design we have $\beta_{\mathrm{ext}}=QE=0.89$ for high-mobility graphene. Thus, for the entangled-photon emission in the case of high-mobility graphene at $\lambda=1.55$ $\mu$m, we obtain $\beta_{\mathrm{pair}}\simeq \beta_{\mathrm{ext}}^{2} = QE^2=0.79$.

\section{Conclusion}
Here we have demonstrated that metallic nanoparticles on an hBN/graphene/SU-8 heterostructure can produce a large Purcell enhancement effect via coupling of SPEs to the supported AGP eigenmodes in the cavity of the heterostructure. We showed that enhancement factors of up to eight orders of magnitude are possible when using hBN thicknesses of 1 nm. We showed that the frequency of these AGPs is determined by the geometry of the cavity, specifically the thickness of the hBN layer and the number of graphene layers intercalated with FeCl$_3$. Importantly, the AGP frequency can also be tuned via a gate voltage applied to the graphene layer which varies the graphene's Fermi energy. This allows for the AGP frequency to be varied with respect to time, enabling the Purcell enhancement to be effectively switched on and off. 
The resulting Purcell enhancement factors are on the order of several million in the mid IR, with radiative enhancements exceeding $10^6$. This enhancement enables high quantum yield for single-photon emission and makes electric quadrupole and two-photon spontaneous emission (2PSE) experimentally accessible. Quadrupole transitions are particularly compelling, as they allow the emitted photons to carry orbital angular momentum. This additional degree of freedom opens new avenues for encoding information, offering significant advantages for quantum communication technologies. By changing the values of $E_F$, $C_S$, and $h$ we can tune the AGP to wavelengths of interest in telecommunications at the expense of a reduction in the enhancement rate. For example, at 1550 nm we demonstrate a Purcell enhancement of $1.76\times10^4$ for dipole transitions,  $1.19\times 10^{7}$ for quadrupole transitions, and $3.1\times 10^{8}$ for 2PSE. Remarkably, we show that using high-mobility graphene it is possible to reach quantum efficiencies of 95\% and 89\% in the mid-IR and near-IR, respectively. In addition, we obtain a quantum efficiency of 79\% for entangled-photon emission with high-mobility graphene at $\lambda=1.55$ $\mu$m. These large Purcell enhancement factors and the high quantum efficiencies at mid-IR and near-IR wavelengths make the metamaterials based on Ag nanoparticles on dielectric/graphene a promising platform for the development of single-photon and entangled-photon sources for use in telecommunications. We also showed that the Purcell enhancement of forbidden transitions can be greatly increased. This enhancement allows for transitions that were previously not detectable by experiment, to be utilized in applications related to quantum information science and engineering.

\section{Methods}

\subsection*{Graphene Material Parameters}
The finite-difference time-domain (FDTD) simulations were performed using the commercial Ansys Lumerical FDTD software. The graphene layer is modeled as a 2D plane with a total conductivity defined as the optical conductivity of graphene with linear dispersion, giving rise to the intraband optical conductivity\cite{Safaei2017,Paudel2017,Safaei2019,SafaeiACS}
\be
\sigma _{\rm intra}(\omega ) = \frac{e^2}{\pi\hbar^2}\frac{2k_BT}{\tau^{-1} - i\omega}\ln \left[ 2\cosh \left( \frac{\varepsilon _F}{2k_BT} \right) \right],
\ee
 which in the case of ${\varepsilon _F} \gg {k_B}T$   is reduced to 
 \be
\sigma_{\rm intra}(\omega) = \frac{e^2}{\pi\hbar^2}\frac{\varepsilon_F}{\tau^{-1} - i\omega }=\frac{2\varepsilon_m\omega_p^2}{\pi\hbar^2(\tau^{-1}-i\omega)},
\label{eq:sigma_intra}
 \ee
 where $\tau$ is determined by impurity scattering and electron-phonon interaction ${\tau ^{ - 1}} = \tau _{imp}^{ - 1} + \tau _{e - ph}^{ - 1}$ .
 Using the mobility $\mu$ of the graphene sheet, it can be presented in the form
 $\tau^{-1}=ev_F^2/(\mu E_F)$, where $v_F=10^6$ m/s is the Fermi velocity in graphene.
 $\omega_p=\sqrt{e^2E_F/2\varepsilon_m}$ is the bulk graphene plasma frequency. We use a graphene mobility of 2000 $cm^2V^{-1}s^{-1}$.
Since the graphene sheets are electronically decoupled from each other by the insulating FeCl$_3$ layers, the optical conductivity of multilayer graphene (MLG) intercalated with FeCl$_3$ is given by\cite{Shabbir2022}
\be
\sigma_{\rm intra}^{\rm MLG-FeCl_3}(\omega) = N\frac{e^2}{\pi\hbar^2}\frac{E_F}{\tau^{-1} - i\omega },
\label{eq:sigma_intra^MLG-RuCl3}
 \ee
 where $N$ is the number of graphene layers. This formula is only valid for excitation energies below the band gap $E_g=1$ eV of FeCl$_3$.
 Since we study AGP resonances at energies between 0.18 eV ($\lambda=7.0$ $\mu$m) and 1.24 eV ($\lambda=1.0$ $\mu$m), we can safely neglect the optical phonons in FeCl$_3$ at 2.7 meV and 7 meV.

To model multiple layers of graphene intercalated with FeCl$_3$, we utilize the conductivity scaling variable in the Lumerical FDTD software. The validity of this approach to model multi-layer graphene is discussed later. The mesh size is set within the dielectric layer, consisting of hBN or hBN with WS$_2$, to be 1 nm in the x and y directions and 0.5 nm in the z direction. The auto shutoff level is set to $10^{-7}$ and the simulation time to 40000 $\mu s$.

The following quantities are useful to understand the methods of extracting the Purcell enhancement factors, the spontaneous decay rates, the normalized spontaneous decay rates, and the quantum efficiency from FDTD simulations.

\subsection*{Purcell Enhancement Factor and Spontaneous Decay Rates}
The emission rate of a quantum emitter, \( \Gamma_{\text{emi}} \), in the presence of an antenna is determined by the excitation rate, \( \Gamma_{\text{exc}} \), and the quantum efficiency, \( QE \). Assuming these effects are uncorrelated, i.e.
\begin{equation}
\Gamma_{\text{emi}} = \Gamma_{\text{exc}} \cdot QE.
\end{equation}
In a homogeneous medium (denoted by superscript ‘0’), we have
\begin{equation}
\frac{\Gamma_{\text{emi}}}{ \Gamma^{0}_{\text{emi}}} = \frac{\Gamma_{\text{exc}}}{ \Gamma^{0}_{\text{exc}}} \frac{QE}{QE^{0}}.
\end{equation}
The enhancement of the excitation rate is given by
\begin{equation}
\frac{\Gamma_{\text{exc}}}{ \Gamma^{0}_{\text{exc}}} = \frac{|\mathbf{E}|^2}{|\mathbf{E}^{0}|^2}.
\end{equation}
The quantum efficiency, \( QE \), is defined as the ratio between radiative decay rates and total decay rates, i.e.
\begin{equation}
QE = \frac{\Gamma_{\text{rad}}}{\Gamma_{\text{rad}} + \Gamma_{\text{nr}} + \Gamma_{\text{loss}}}.
\end{equation}

\subsection*{Normalized Spontaneous Decay Rates}

Absolute decay rates \( \Gamma_{\text{rad}} \), \( \Gamma^{0}_{\text{rad}} \), \( \Gamma_{\text{loss}} \) cannot be directly calculated, but their normalized forms can by using the ratio
\begin{equation}
\frac{\Gamma_{\text{dip}}}{\Gamma^{0}_{\text{rad}}} = \frac{P_{\text{dip}}}{P^{0}_{\text{rad}}}.
\end{equation}
The Purcell enhancement factor is
\begin{equation}
\calF=\frac{\Gamma_{\text{dip}}}{\Gamma^{0}_{\text{rad}}}.
\end{equation}
The normalized spontaneous decay rates are
\be
\frac{\Gamma_{\text{rad}}}{\Gamma^{0}_{\text{rad}}} = \frac{P_{\text{rad}}}{P^0_{\text{rad}}},
\quad
\frac{\Gamma_{\text{loss}}}{\Gamma^{0}_{\text{rad}}} = \frac{P_{\text{abs}}}{P^0_{\text{rad}}}.
\ee

\subsection*{Quantum Efficiency}

For a high intrinsic quantum efficiency (assuming \( QE^{0} = 1 \) and \( \Gamma^0_{\text{nr}} = 0 \)) we have
\begin{equation}
\frac{QE}{QE^{0}} = QE = \frac{\Gamma_{\text{rad}}}{\Gamma_{\text{rad}} + \Gamma_{\text{loss}}}.
\end{equation}
This is the ratio of power radiated to the sum of radiated and absorbed power. The QE of the device is plotted for both wavelengths of interest in Fig.~\ref{fig:QE}.

\section*{Acknowledgment}
M.A.K., M.N.L., and D.R.E. acknowledge support by the Air Force Office of Scientific Research (AFOSR) under award no. FA9550-23-1-0472.

\bibliographystyle{achemso}
\bibliography{bibliography}

\end{document}